# Interactive Execution Monitoring of Agent Teams

**David E. Wilkins**                                                      WILKINS@AI.SRI.COM
**Thomas J. Lee**                                                          TOMLEE@AI.SRI.COM
**Pauline Berry**                                                            BERRY@AI.SRI.COM
*Artificial Intelligence Center, SRI International*
*333 Ravenswood Ave., Menlo Park, CA 94025 USA*

## Abstract

There is an increasing need for automated support for humans monitoring the activity of distributed teams of cooperating agents, both human and machine. We characterize the domain-independent challenges posed by this problem, and describe how properties of domains influence the challenges and their solutions. We will concentrate on dynamic, data-rich domains where humans are ultimately responsible for team behavior. Thus, the automated aid should interactively support effective and timely decision making by the human. We present a domain-independent categorization of the types of alerts a plan-based monitoring system might issue to a user, where each type generally requires different monitoring techniques. We describe a monitoring framework for integrating many domain-specific and task-specific monitoring techniques and then using the concept of *value of an alert* to avoid operator overload.

We use this framework to describe an execution monitoring approach we have used to implement Execution Assistants (EAs) in two different dynamic, data-rich, real-world domains to assist a human in monitoring team behavior. One domain (Army small unit operations) has hundreds of mobile, geographically distributed agents, a combination of humans, robots, and vehicles. The other domain (teams of unmanned ground and air vehicles) has a handful of cooperating robots. Both domains involve unpredictable adversaries in the vicinity. Our approach customizes monitoring behavior for each specific task, plan, and situation, as well as for user preferences. Our EAs alert the human controller when reported events threaten plan execution or physically threaten team members. Alerts were generated in a timely manner without inundating the user with too many alerts (less than 10% of alerts are unwanted, as judged by domain experts).

## 1. Introduction

As automation and reliable, high-bandwidth communication networks become more common, humans are increasingly responsible for monitoring and controlling the activity of distributed teams of cooperating agents, both human and machine. Such control decisions in many realistic domains are complex, and require human experience and judgment. Our vision is that human decision makers will be able to perform more important tasks than continuously monitoring incoming information by relying on an automated execution aid to alert them when significant new information warrants their attention. We are primarily interested in domains requiring human control and will describe two such domains. However, the majority of our techniques and analysis also apply to completely automated execution monitoring. In fact, in one of our domains we both interact with a human controller and autonomously adjust robot behavior and plans.

To rapidly make effective control decisions for distributed agent teams, the human needs automated support, for several reasons. First, inexpensive sensors and reliable, high-bandwidth communication networks provide large volumes of pertinent data arriving from sensors, team members, and other sources. Without automated support, the human cannot cope with the volume of incoming information. Second, plans that coordinate the activity of several team members, as many as several hundred in our first domain, can become too complex to monitor without automated help. Third, we are addressing domains that are dynamic, sometimes requiring responses in a few seconds or less. Fourth, the automated team members (robots) are complex, with different failure modes and recovery procedures, and automated support for controlling them is often essential. All these challenges are magnified as the tempo of the decision cycle increases or the user becomes stressed. Thus, domains with the above properties require an interactive, automated assistant to support humans in monitoring incoming information and controlling agent teams.

We will concentrate on dynamic, data-rich domains where humans are ultimately responsible for team behavior. Realistic domains often have adversaries to overcome. These may range from fairly benign forces of nature that introduce uncertainty, to intelligent adversaries that are trying to actively thwart plans. An automated execution assistant should interactively support effective and timely decision making by the human, and interact with the human to take advantage of knowledge the human possesses that is not explicitly modeled in the machine. Ideally, an execution assistant would allow its human user to, among other things:

- Guide the system with minimal effort

- Focus on external events, assuming the system will alert the user when human attention is desirable

- Understand, evaluate, and modify the plans/actions

- Understand why and why not for each action or decision taken/recommended/rejected by the system

- Have constant multimodal feedback

- Recommend actions and decisions that violate constraints when warranted

One key idea is that rich plan representations allow the execution aid to share context with users, so both understand the semantics of plans and requests. Understanding the plan is the key to helping the user deal with the possible information glut created by advanced information systems. The execution aid uses the plan to filter, interpret, and react to the large volume of incoming information, and can alert the user appropriately when events threaten the plan or the user's physical existence.

Once the user develops trust in the execution aid, there will be a reduction of the need for human monitoring of the display of the information system, while simultaneously increasing the amount of relevant information monitored because the aid analyzes every piece of incoming data. Relying on alerts from an automated aid allows the human to pay attention to more important tasks than monitoring incoming data, attending to the display only when alerted by the execution aid.

In the next section, we characterize the domain-independent challenges posed by this problem, concentrating those that are unique to interactive execution aids in dynamic domains with distributed teams of cooperating agents. Then, we describe how properties of various domains influence these challenges and their solutions. In Section 4, we present a domain-independent categorization of the types of alerts a plan-based monitoring system might issue to a user. Next, we describe the concept of value of information and alerts that is key to reducing unwanted alerts (alarms). Sections 6 and 7 describe the Execution Assistants we implemented in the small unit operations and robotics domains, respectively. Sections 6.8 and 7.5 contain the results of evaluations performed in each domain. Finally, we discuss related work and present our conclusions.

## 2. Interactive Monitoring Challenges

There has been great interest in plan generation algorithms, but less work on using plans to dynamically control execution. Much execution monitoring work describes monitors in specific domains, so we first characterize the domain-independent challenges of monitoring agent teams.

There are several "universal" challenges of execution monitoring that are not particular to dynamic, data-rich domains or interactive monitoring. These issues should be part of a monitoring ontology and are addressed in our EAs, but we do not stress them in our discussion as they are discussed elsewhere (Kaminka, Pynadath, & Tambe, 2001; Jonsson, Morris, Muscettola, & Rajan, 2000; Muscettola, Nayak, Pell, & Williams, 1998; Myers, 1999; Wilkins, Myers, Lowrance, & Wesley, 1995; Coiera, 1993; Durfee, Huber, Kurnow, & Lee, 1997). The issues include the following:

- Sensitivity of the monitor — its ability to detect problems or meet requirements. The system must remain reactive to incoming data while performing monitoring tasks.

- Temporal reasoning and temporal sensitivity. Execution takes place over time and plans specify future actions, thus making temporal reasoning central.

- Concurrent temporal processes. Multiple tasks or agents may be executing concurrently.

- Synchronization between agents. An execution assistant must get the right information to the right team members at the right time to support the cooperative activity specified in the plan. In some domains, this may require doing plan recognition on other team members (Kaminka et al., 2001).

- False and redundant alarms. Unwanted alarms are ubiquitous in data-rich domains such as medicine (Koski, Makivirta, Sukuvaara, & Kari, 1990; Tsien, 1997) and the domains described in this paper.

- Combining event-driven and goal-driven behavior. The execution assistant must respond to unfolding events with acceptable latency while concurrently invoking actions that will continue execution of the (perhaps modified) plan and satisfy user requests. Goal-driven tasks include responses to events, such as generating modified, new, or contingency plans, and invoking standard operating procedures.

- Adversarial reasoning, including plan and pattern recognition. Many real-world domains have adversaries and their activity must be closely monitored.

We are concerned with execution monitoring of agent teams, where team members may be any combination of humans and/or machines. We concentrate on the challenges that are unique to interactive execution aids in dynamic domains, and categorize these challenges into the following four categories.

**Adaptivity.** The output of an execution assistant must meet human requirements and preferences for monitoring behavior, providing high-value alerts and suggestions. As in all execution monitoring, sensitivity is crucial, but in interactive monitoring the sensitivity of the monitor must also be adaptable. In addition to adapting to user preferences, the analysis done by an execution assistant and its level of autonomy must be adjustable to operational tempo and incoming data rate. The system should ideally adapt its output to the user's capabilities and cognitive load.

**Plan and situation-specific monitoring.** Coordinating the activities of many teams members requires a plan shared by the team. We will assume that plans contain partial orders of *tasks* for each team member, as well as any necessary coordinating instructions and commitments (Grosz & Kraus, 1999). The plan representation also encodes some of the expected outcomes (effects) of plan execution, so that execution aids can detect deviations. The analysis done by an execution assistant and any suggested responses must depend on the plan and situation to be effective, because events often cause a problem for some plans but not for others. We found that monitoring algorithms must often be tailored to the specific tasks that compose plans. To facilitate interaction, the plan representations must be understandable by both humans and the system, although the human might be aided by multiple plan views of the internal representation in a user-friendly interface.

**Reactivity.** Any execution monitor must react to events and uncertainty introduced by the environment. In dynamic, data-rich domains, particular care must be taken to ensure that the system remains reactive with high rates of incoming information and fast decision cycles. Resources are not generally available to perform all desired analyses for every input — for example, projecting future problems with multiple simulation runs or searching for better plans may be computationally expensive. There are often no obvious boundaries to the types of support an execution aid might provide in a real-world domain. Therefore, a balance must be struck between the capabilities provided and resources used. A few examples show the types of issues that arise in practice. In our first domain, only coarse terrain reasoning was used, as projections using fine-grained terrain data were computationally expensive. In our robot domain, we had to adjust the time quanta assigned to processes by the scheduler so that our monitoring processes were executed at least every second. Finally, in domains with dangerous or intelligent adversaries, reacting to their detected activity becomes a high priority. There has been considerable research on guaranteeing real-time response (Ash, Gold, Seiver, & Hayes-Roth, 1993; Mouaddib & Zilberstein, 1995), but the tradeoffs are generally different in every application and are usually a critical aspect of the design of an execution assistant.

**High-value, user-appropriate alerts.** Alerting on every occurrence of a monitored condition that is possibly a problem is relatively easy; however, the user would quickly ignore any assistant that gave so many alerts. The challenge is to not give false alarms and to not inundate the user with unwanted or redundant alerts. The system must estimate the utility of information and alerts to the user, give only high-value alerts, and present the alerts in a manner appropriate to their value and the user's cognitive state. We found that a common challenge is to avoid cascading alerts as events get progressively further away from expectations along any of a number of dimensions (such as time, space, and resource availability). Another challenge that we will not discuss in depth is aggregating lower-level data (e.g., sensor fusion), which can reduce the number of alerts by consolidating inputs. Estimates of the value of alerts can be used to adjust alerting behavior to the user's cognitive load.

Interactive alerting during execution naturally leads to the equally important and challenging topic of human directing of responses and plan modifications. Our monitoring technologies have been used in continuous planning frameworks (Wilkins et al., 1995; Myers, 1999), but we will limit the scope of this paper to interactive alerting. We briefly mention some ongoing research on this topic that we either are using or plan to use in conjunction with our execution aids.

Agent systems that interact with humans are an active area of research, and the issues are discussed in the literature (Myers & Morley, 2001; Ferguson & Allen, 1998; Schreckenghost & et al., 2001). Myers and Morley (2001), for example, describe the Taskable Reactive Agent Communities (TRAC) framework that supports human supervisors in directing agent teams. They address topics such as adjustable agent autonomy, permission requirements, consultation requirements, and the ability to communicate strategy preferences as guidance. TRAC is complementary to the execution monitoring described in this paper.

Another active research area that fits naturally with our execution monitoring approach is theories of collaboration. In fact, we use the SharedPlans theory of collaboration (Grosz & Kraus, 1999) in our second domain (Ortiz & Hsu, 2002) to direct agents in conjunction with the execution monitor. This theory models the elements of working together in a team as well as the levels of partial information associated with states of an evolving shared plan. Central to the theory of SharedPlans is the notion that agents should be committed to providing helpful support to team members. Within the theory, this notion of helpful behavior has been formally defined (Ortiz, 1999). The work on collaboration is complimentary with our monitoring approach, but will not be discussed in detail.

## 3. Monitoring Approach Determined by Domain Features

The domain features and monitoring challenges with which we are concerned are common in many domains in addition to robot teams and small unit operations (SUO). For example, they occur in the monitoring of spacecraft (Bonasso, Kortenkamp, & Whitney, 1997; Muscettola et al., 1998) and monitoring in medicine (Coiera, 1993) for ICU patients or for anesthesia. These domains are also data rich — medical clinicians have "difficulty in using the vast amount of information that can be presented to them on current monitoring systems" (Weigner & Englund, 1990; Coiera, 1993). In particular, the problem of flooding human users with false or redundant alarms is ubiquitous in medical monitoring (Koski et al., 1990; Tsien, 1997). One study found that 86% of alarms in a pediatric ICU were false alarms (Tsien & Fackler, 1997). False alarms distract humans from more important tasks. Such a false alarm rate would most likely make the monitor useless in fast-paced operations. Research in these domains has concentrated on automated monitoring, with little or no emphasis on interactive monitoring.

While the challenges described in the previous section apply to all interactive, dynamic domains, the properties of individual domains influence their solutions. One brief case study shows how the features of the communication system and the use of legacy agents can indicate a different monitoring approach for two similar problems. Kaminka et al. (2001) address a problem similar to ours: many geographically distributed team members with a coordinating plan in a dynamic environment. They use an approach based on applying plan-recognition techniques to the observable actions of team members, rather than communicating state information among team members, which they refer to as *report-based monitoring*.

They list four problems with report-based monitoring (Kaminka et al., 2001): (1) intrusive modifications are required to legacy agents to report state, (2) the necessary state information changes with the monitoring task, (3) the monitored agents and the communication lines have heavy computational and bandwidth burdens, and (4) it assumes completely reliable and secure communication between the team members. They say that (1) is their main concern, with (3) being next most important.

| Plan constraint violated |
| Policy constraint violated |
| New opportunity detected |
| Adversarial activity detected |
| Constraint violation, opportunity, or adversarial activity projected |
| Contingency plan suggested |
| System problem detected |
| Reporting requirement triggered |

Figure 1: Top-level categories in alert ontology.

In both of our domains, we use report-based monitoring. Our agents already report their state or can easily be modified to do so, for example, by attaching Global Positioning (GPS) devices. Our monitoring tasks can be performed using the reports already available, although one can imagine adding further functionality that would change the reporting requirements. In our first domain, reports are distributed by the Situation Awareness and Information Management (SAIM) system on a high-bandwidth network. SAIM uses novel peer-to-peer (P2P) dissemination algorithms and forward fusion of sensor reports, greatly reducing bandwidth requirements. P2P is fault tolerant, allowing any node to be a server. Dissemination is based on an agent's current task, geographic location, and relationship in the hierarchical organization of team members.

In summary, report-based monitoring works in our domains because we rely less on unmodifiable legacy agents, have more reliable communications, and have enough bandwidth available with our network and dissemination algorithms. Kaminka's approach provides more automated support, but we must address the problem of modeling the value of information to the user. If Kaminka's system was extended to interact with humans, we believe our alert ontology and techniques for avoiding operator overload would be applicable, whether alerts come from sources based on plan-recognition or from reports. Because we rely on humans as being ultimately responsible for team behavior, we do not require as much state information nor complete reliability in communication. Unreliable communication will degrade monitoring performance, but the human decision maker must take missing inputs into account when making a decision. The execution assistant can monitor communications and alert the human to possible communications problems.

## 4. Types of Alerts

Alerts are used to focus the user's attention on an aspect of the situation that the execution aid has determined to be of high value. We discuss the problem of determining the value of information and alerts in later sections, which determines whether and how an alert is presented. An alert may indicate that a response is required, or may just be informative. Many different types of alerts can be given, and it is useful to categorize alerts, thus providing the beginning of a reusable, domain-independent ontology for execution monitoring.

Figure 1 shows the top-level categories for alerts that we identified by starting with a super-set of the categories we found useful in our two domains and then generalizing them to cover a broad range of domains. It is assumed that execution is directed by a plan that is shared by the team. These categories generally require different monitoring techniques and different responses to detected problems. For example, adversarial activity could have been a subclass of other relevant classes, but it requires different monitoring techniques. The friendly location data is precise (within the error of GPS) and trustworthy, while adversarial data comes from fusion engines running on data from sensor networks. The adversarial data is highly uncertain, may come at significantly different rates, and generally will have different algorithms for determining the value of information, as adversarial entities are actively trying to thwart your plan and perhaps are trying to kill you.

The top-level categories in our ontology generally differ along the following dimensions that are important to monitoring:

- Properties of data sources (such as reliability and uncertainty).

- Rates of incoming data

- Method of acquiring data (such as receiving messages, pulling data from databases, doing plan recognition)

- Monitoring algorithms, including tradeoff of complexity of analysis with reactivity

- Desired responses to alerts

- Value of information algorithms

The different monitoring techniques for each category are often domain specific, and can even be task specific in some cases, adapting the monitoring as tasks in the plan are executed. Our monitoring framework integrates these various techniques and then uses the concept of *value of an alert* to control interaction with the user.

We briefly discuss each of the top-level categories. We have not provided the next lower level of the ontology because the space of possibilities is large, with domain-specific concerns important. For example, adversarial alerts could include subclasses for fixed or mobile adversaries, for size and capabilities of the adversarial team, for an alliance or tightly coordinated adversarial team, for adversarial intent or plan, and so forth. Later in the paper, we describe how alerts given by our implemented execution assistants (EAs) fit into these categories.

**Plan constraints.** Plans provide most of the expectations of how execution should proceed, so this category has the richest set of alerts. A fairly large hierarchical ontology could be produced to describe different types of alerts on plan constraints. Gil and Blythe (1999) present a domain-independent ontology for representing plans and plan evaluations. Each concept in their evaluation ontology could be a source of an alert when the evaluation becomes sufficiently important to the user. Plans in real-world domains are often hierarchical, so constraints from different levels or layers may be violated. It may be desirable to customize alerts based on the hierarchical level of the plan constraint in question. To indicate the range of possible alerts in this category, we list a few common examples:

- A coordinating team member (or the agent) is out of position or late.

- The effects of the agent's (or a team member's) actions were not achieved as expected.

- A team member has retracted a commitment to perform a certain task, requiring a reallocation of tasks or resources.

- Conditions required by the plan are not true when expected.

- Resources used by the plan are not available or degraded.

**Policy constraints.** Most real-world domains have persistent constraints such as policies or rules of engagement that must not be violated. While these could be considered as part of the plan by representing them as maintenance conditions that extend over the entire plan, they are significantly different in practice and are often monitored by different techniques, because they may require additional domain knowledge or specialized monitoring algorithms, which must be invoked efficiently. For example, in our domains, we never want our human team members to be killed or our robots destroyed. Therefore, we monitor the physical safety of our agents at all times and give alerts to the user when some agent is in danger. Dangers from adversarial agents are covered in their own category. However, the system should also alert the user to threats from team members (fratricide) and from the local agent's own actions (e.g., a robot's battery running low).

**New opportunities.** Even though the current plan can still be executed without change, it may be possible to generate a better plan for the current situation as new opportunities arise. Determining if an execution-time update to the world state permits a more desirable plan is a difficult problem in general, similar to generating a new plan for the new situation. However, in real-world domains, there are often methods for detecting new opportunities that indicate a plan revision might be cost effective. For example, certain key features (such as "pop-up targets" in military domains) can represent new opportunities, and there are often encoded *standard operating procedures* (SOPs) that can be invoked when triggered by the current situation to improve the plan and/or react to events. Because our monitoring is interactive, we can avoid the difficult decision of whether to search for a better plan by alerting the user of high-value opportunities and relying on the user to judge the best response.

**Adversarial activity.** This category assumes that our team members are operating in environments with adversaries that are trying to actively thwart team plans. When adversaries are dangerous (e.g., worthy human opponents), reacting to their detected activity becomes a top priority and, in our experience, merits customized monitoring algorithms. Recognizing immediate threats to a team member's physical existence or to the accomplishment of the plan is obviously important. In addition, information that allows the human to discern patterns or recognize the opponent's plan or intent is valuable. Our EAs recognize physical threats and adversarial activity not expected by the plan, but do not currently perform automated plan or intent recognition on data about adversaries. Both automated plan recognition (Kaminka et al., 2001) and inference of adversarial intent (Franke, Brown, Bell, & Mendenhall, 2000; Bell, Jr., & Brown, 2002) are active areas of research. If algorithms are developed that reliably recognize adversarial plans or intent while using acceptable computational resources, they could easily be invoked within our monitoring framework.

**Projections.** Even though the current plan can still be executed without change for the time being, it may be possible to predict that a future failure of plan or global constraints will occur, with varying degrees of certainty. For example, suppose the plan requires a robot to move to location X by time T, but the robot is getting progressively more behind schedule or more off course. At some point before T, the system can predict with acceptable certainty that this location constraint will be violated and alert the user, who may revise the plan. In addition, new opportunities and probable adversarial activity could be projected. Projection/simulation algorithms can be computationally expensive, so the execution monitor must adjust its calculation of projections to match available resources and constraints.

**Contingency plans.** The plan may specify contingency plans or subplans, which are to be invoked when certain, specified conditions arise. The execution monitor should monitor these conditions and alert the user when a contingency plan has been triggered. The system can also notify all team members automatically if the user decides to switch execution to a contingency plan. Another desirable alert in some domains might be a suggestion by the system that new contingency plans should be generated for certain situations as events unfold in an unexpected manner. Our EAs monitor the triggering of contingencies but do not suggest their generation.

**System problems.** Depending on the domain, the user may want to be alerted of problems with incoming data streams or in the functioning of the execution assistant itself. For example, if no data is arriving from the sensors, or over the network from other team members, this may be crucial to helping the user interpret the situation and system alerts.

**Reporting requirements.** One of our basic assumptions is that the human user has experience and knowledge that are not modeled within the system. Therefore, the system cannot always recognize how a new piece of information will affect plan execution. Some information that does not trigger the above alerts might still be valuable to the user. The system is given reporting requirements that allow it to recognize such information. One generally useful reporting requirement would be execution status, so the user can quickly determine that execution is proceeding as planned. Reporting requirements may take any number of forms, as appropriate to the domain. The comments about recognizing new opportunities apply here — domains might specify requirements as SOPs, key features, declarative statements, or heuristic algorithms. Several things fall under this category, such as information that reduces uncertainty and/or indicates that the plan is executing as expected. As another example, a robot might be told to immediately report any murder or fire it witnesses while executing its planned tasks.

## 5. Value of Information and Alerts

Algorithms that alert on constraint violations and threats in a straightforward manner inundate the user in dynamic domains. Unwanted alerts were a problem in both our domains and in many other domains as well, such as medical monitoring (Koski et al., 1990). An aid that gives alerts every second will quickly be discarded by the user in stressful situations (if not immediately). To be useful, an execution aid must produce high-value, user-appropriate alerts. Alerts and their presentation may also have to be adjusted to the situation, including the user's cognitive state (or the computational state of a software agent). For example, in high-stress situations, tolerances could be increased or certain types of alerts might be ignored or postponed. In this section, we provide a conceptual framework for the alerting algorithms in our monitoring framework and our domain-specific EAs.

Our approach is grounded in the concept of determining the value of an alert. First, the system must estimate the value of new information to the user. Information theory derives from communication theory and the work by Shannon (1948). In this theory, the value of information refers to the reduction in uncertainty resulting from the receipt of a message, and not to the meaning that the message (or the uncertainty reduction) has to the receiver (Weinberger, 2002). We use the term *value of information* (VOI) in a different sense, namely, the pragmatic import the information has relative to its receiver. (Of course, the reduction of uncertainty often has pragmatic import.) Like Weinberger (2002), we assume that the practical value of information derives from its usefulness in making informed decisions.

However, alerting the user to all valuable information could have a negative impact in certain situations, such as when the alert distracts the user from more important tasks, or when too many alerts overwhelm the user. We therefore introduce the concept of *value of an alert* (VOA), which is the pragmatic import (for making informed decisions) of taking an action to focus the user's attention on a piece of information. VOA takes VOI into account but weighs it against the costs and benefits of interrupting the user. If the user is busy doing something significantly more important, then issuing an alert might not be valuable, even when VOI is high. VOA must generally estimate the user's cognitive state and current activities. VOA will generally determine the modality and other qualities of alert presentation (e.g., whether one should flash red text on a computer display or issue a loud audible warning).

VOI and VOA are highly correlated in most situations, and most general comments about VOI apply to VOA as well. However, VOA may be low while VOI is high if the user is highly stressed or preoccupied with more important tasks. It is also possible to have a high VOA and low VOI. For example, mission-specific monitors might alert the user to information that has been known for some time (and thus has little or no value as information) because the information is crucial to an upcoming decision and the user may have forgotten it, or may be behaving in a way that indicates a lack of awareness.

Weinberger gives a quantitative definition of pragmatic information, assuming a finite set of alternatives that lead to well-defined outcomes, each of which has some value to the decision maker. In realistic domains like ours, alternatives and outcomes are not precisely defined. Furthermore, information and decision theories (including Weinberger's) assume that the decision maker is aware of (or has processed) previous information and can devote sufficient resources to analyzing the current information. Under such assumptions of unlimited processing power, VOA and VOI are the same. In most realistic domains, these assumptions do not hold. Humans are resource bounded and, during fast-paced operations, alerts and information may be ignored and the user may not realize the implications of information on a complex plan that coordinates many team members.

## 5.1 Estimating VOI and VOA

In interactive, dynamic, real-world domains like SUO, we cannot model all alternatives, their payoffs, nor all the other knowledge and probabilities required with enough precision to compute the "theoretical" VOI and VOA. Much knowledge about VOI resides only with human experts, and even they might have different preferences or opinions about VOI. For example, in the SUO domain, the user might be concerned about the public-relations effects of how the plan execution is reported in the international media. It is precisely because humans have knowledge not modeled in the system that we want our execution assistants to be interactive. In such realistic domains, there are generally no obvious boundaries to the types of support the system should provide, and no precisely defined evaluation functions or payoff matrixes. Thus, Weinberger's theory and formal techniques for computing the value of information (Athey & Levin, 2001) cannot be applied. Horty and Pollack (2001) develop some of the foundations for a theory of rational choice that involves estimating the cost of decisions in the context of plans. Their approach comes closer to addressing our concerns. However, determining costs and utilities of actions will continue to require human judgment in many domains, especially if human lives are being put at risk.

Therefore, we developed algorithms that heuristically estimate VOI using domain knowledge, although quantitative VOI functions can easily be used in our framework. The inputs to our al-gorithms are described in Section 5.3. These domain-specific algorithms are, and must be, easily customized and tuned for user preferences, as well as the situation. They are invoked in domain-independent ways for a variety of purposes by the monitoring framework, and were developed with feedback from domain experts. We believe it is feasible to use machine-learning techniques to replace or supplement hand-coded heuristics for VOI/VOA estimation and/or the user preferences which affect it, but this was not explored.

VOI and VOA are computed qualitatively in our domains, using several domain-specific quantitative measures in the qualitative reasoning process. Issuing an alert is a discrete event, and generally there are a small number of options for presenting an alert. Therefore, estimating VOA is primarily a problem of categorizing the potential alert into a small number of alert presentation types or modalities. We need to determine when the VOA crosses thresholds (defined by the VOI/VOA specification) indicating, for example, that it is valuable to issue an alert, or that the alert should be issued as high-priority. In our framework, the thresholds are customizable by the user and can be mission specific, so they can change automatically as different missions in the plan are executed. The VOI algorithms also determine what information to include in an alert.

Different alert presentations are handled by assigning a qualitative *priority* to each alert. For example, our SUO EA divides alerts by VOA into four equivalence classes for levels of priority, which were already defined in the SUO domain. Each priority is presented differently to the user, from using different modalities to simply using different colors or sizes of text or graphics. Currently, we use three priority levels in the robotics domain, but may add more in the future as collaborating team members make more use of the EA. These priority levels can be used to adjust alerting behavior to the user's cognitive load. For example, during fast-paced operations, only the highest-priority alerts could be presented.

There are several reasons for preferring qualitative reasoning, and we draw on Forbus's work in describing the advantages (Donlon & Forbus, 1999; Forbus, 2002). Qualitative models fit perfectly with making decisions, which are discrete events, and effectively divide continuous properties at their important transitions. Thus, changes in qualitative value generally indicate important changes in the underlying situation. Qualitative models also facilitate communication because they are built on the reasoning of human experts and thus are similar to people's understanding. For example, the priority levels used in our VOA algorithms have long been named and defined in the military. Qualitative reasoning is important as a framework for integrating the results of various qualitative computations in a way humans can understand. Finally, the precision of quantitative models can be a serious weakness if the underlying models do not accurately reflect the real-world situation. Precise data can lead to precise but incorrect results in a low-accuracy model, and the precise results can lead to a false sense of security.

These advantages of qualitative reasoning are apparent in both common sense and military reasoning. Common sense reasoning about continuous quantities is often done qualitatively. The continuous value is of interest only when a different action or decision is required. For example, you can ignore your fuel gauge when driving once you have decided whether or not you must refuel before reaching your destination. In addition to the priorities already mentioned, the military quantizes many continuous properties used to describe terrain in ways that are relevant to military operations, creating phase lines, decision points, named areas of interest, key terrain avenues of approach, and so forth. The SUO EA incorporates these quantizations to reason about terrain's influence on VOI and VOA and to effectively communicate information in alerts, just as the military has used them for years to facilitate communication, collaboration, and decision making.

## 5.2 Properties of VOI and VOA

VOI and VOA in our dynamic domain depends primarily on whether the information will influence decisions/responses. The execution aid must also ensure human awareness of high-value data to support decisions only the human user can make. Thus, the system must estimate or model what the human needs to know (e.g., by specifying reporting requirements), even if the system cannot predict how the information might influence a decision. For example, an emerging adversarial or friendly pattern might be crucial. If the system does not have a human-level ability to recognize plans and patterns, then it should ensure the human decision maker is aware of the relevant data.

One obvious but important property of VOI is that it is zero if the user is already aware of the information. Another property is that information indicating that plan execution is proceeding according to plan can be valuable, because it influences the decision to continue as planned. The value of such *confirming* information depends on the features of the domain — such information will be more valuable in domains with high uncertainty and active adversaries.

Another feature that may be useful in certain domains is classifying the responses suggested by a piece of information or an alert. For example, any new report may require a significant plan modification, a minor plan modification, the invocation of a contingency plan, the application of a standard operating procedure (SOP), or the identification of a new opportunity. However, the type of response does not necessarily correlate with VOI, as a minor plan modification might be life saving, while a major modification might simply reduce resource usage by ten percent. The distinction is important because the simpler responses can more likely be handled in an automated fashion, thus reducing the need to involve the user.

Determining what information to present in an alert requires addressing human factors. Initially, it is important to present an alert concisely so the human can determine its import at a glance, and assess whether to divert his or her attention from other tasks. In our EAs, the user can drill down for more detailed information on any alert in order to assess the situation more accurately. Finally, some domains may have concerns other than making informed decisions. For example, the emotional state of the user or recording data of scientific value might be beneficial. In particular, if the concern is analyzing or debugging system performance rather than making good execution decisions, a different VOI estimator can be used to provide alerts about system behavior.

## 5.3 VOI and VOA Criteria

As described above, the VOI and VOA algorithms will generally be heuristic, domain-specific, and user customizable. Here we identify most of the inputs that will be applicable to most interactive, dynamic domains. We started with a superset of the VOI criteria we found useful in our two domains and then generalized them to be domain independent. (The properties of the user listed below are estimates from system models of the user, as the user's mental state is not accessible.)

- The plan

- Policies

- User's awareness of current situation

- System's view of current situation

- User's cognitive load

- Resources, especially time, available for analysis or response

- Information about adversarial agents

- Characterization of uncertainty

- Age of information and age of user's awareness

- Source of information

The plan provides several VOI criteria: the plan may provide explicit and implicit decision points, high-value places, times, team members, and so forth. The value of a task, constraint, adversarial action, or team member is often determined by the plan structure and plan annotations. The tasks in the plan can invoke task-specific VOI algorithms within our monitoring framework, as described in Section 6. Domain policies (or specialized reasoners that implement them) and reporting requirements should provide the knowledge necessary to determine the value of alerts about various types of constraint violations and reports. For example, in our domains, we monitor the physical safety of our agents. Alerts on life-threatening situations have the highest priority.

We noted that VOI tends to zero to the extent the user is already aware of the information. Thus, determining VOI must access the current view of the situation to determine if arriving reports offer new information or simply confirm the existing view. In data-rich domains, we assume that the execution aid may have a more detailed description of the situation than the user (for the aspects of the situation that are described by incoming data), because the user may be performing other tasks and monitoring the situation only when he is alerted by the EA. Therefore, the value of alerting the user will depend on how much the new information differs from the user's last situation update, even if the system has more recent data that differs only slightly from the new information.

Ideally, we would like to model the user's cognitive load, and give lower values to noncritical alerts when the user is consumed with addressing more critical aspects of the situation. Similarly, we do not want to overload the system's computational resources or ability to remain reactive, so the value of certain information may depend on the time or resources available to analyze it.

When determining the value of information about adversaries, it is often useful to compare developing patterns to any information about the adversary's plans or tendencies, which could be obtained from human intelligence analysts or generated by plan-recognition or pattern-matching algorithms. As mentioned above, information that reduces uncertainty is valuable in domains with high uncertainty and active adversaries. VOI can be estimated if we have a characterization of the uncertainty present in our current view of the situation.

The age of information is also a factor in VOI — outdated reports may have zero value if newer information has already arrived. When modeling the user's awareness, elapsed time is a factor. The user will be aware of alerts issued in the last few minutes, but may no longer be aware of something that was brought to her attention yesterday or last week. Thus, the value of a proposed new alert may increase with elapsed time since a similar alert was issued.

When a variety of sources of information exists, the source is a factor in VOI. Often, different information sources have inherently different levels of certainty, authority, or importance. For example, the SUO EA accepts reports from both human observers and automated sensors. An EA with such inputs might want to weigh human observations differently depending on the human and the situation. In later sections on our implemented EAs, we describe our domain-specific VOI/VOA algorithms, which have inputs corresponding to the inputs listed above.

## 6. Implementing Execution Monitors – Small Unit Operations

We have developed an execution-monitoring framework that can easily be adapted to produce interactive monitors for agent teams in dynamic domains. To support this claim, we describe two dynamic, data-rich, real-world domains and the Execution Assistants (EAs) we have implemented using our framework. Our first domain, Army small unit operations (SUO), has hundreds of mobile, geographically distributed agents, which are a combination of humans, robots, and vehicles. The other domain, UV-Robotics (Ortiz, Agno, Berry, & Vincent, 2002), is described in Section 7 and has teams composed of a handful of cooperating, unmanned ground and air vehicles (UGVs and UAVs) and a human controller. Both domains involve unpredictable adversaries in the vicinity of the team members.

We originally developed our monitoring framework for the SUO domain using several person-months of effort, although the majority of the effort was in knowledge acquisition and modeling. The SUO monitoring framework, described below, was designed to be modular and to support the easy insertion of domain-specific (and user-customized) system components, such as task models, monitoring algorithms, and value-of-information estimators. Our design was validated when we implemented a complex execution monitor in the UV-Robotics domain in about one person-week (as described in Section 7.2). The UV EA uses the same plan representation and basic architecture as the SUO EA, but the inputs are different as are the tasks and the monitoring algorithms that respond to the inputs and generate alerts.

The majority of our framework also applies to completely automated execution monitoring as demonstrated by the UV EA. A UV EA runs on each robot in the team and is used to autonomously adjust the robot control by blending desired behaviors and automatically revising plans during execution. The UV EA also provides alerts to any human controller who is monitoring the robots. While the framework described in this section is general, we follow it with some domain-specific details which clarify the concepts and tradeoffs. These details may not be of interest to all readers.

### 6.1 SUO Problem Description

Small unit operations in the military involve hundreds of mobile, geographically distributed soldiers and vehicles cooperatively executing fast-paced actions against an unpredictable adversary. Computational support is bandwidth restricted and must use lightweight and portable devices. Currently, the planning decisions are all made by humans, and the plans are not machine understandable.

We implemented the SUO EA as part of a larger system: the Situation Awareness and Information Management (SAIM) system, which distributes timely, consistent situation data to all friendly agents. SAIM uses new technologies to demonstrate a new concept of automated support (described below) in the SUO domain. We assume many small teams of agents (human, vehicles, and eventually robots), separated and dispersed throughout a large space, operating in concert to achieve goals. We assume that each agent has equipment providing robust geolocation (GPS), computing, and communication capabilities. SAIM also assumes an unpredictable adversary, fast-paced action, and a rich population of sensors controlled by cooperating team members.

The key innovations of SAIM, in addition to the EA, are a self-organizing peer-to-peer information architecture and forward fusion and tracking. Fusion of information and tracking is distributed and done close to the source to minimize latency, bandwidth requirements, and ambiguity. Adjudication maintains consistency of distributed databases. The information architecture supports ad hoc information dissemination based on multicast groups centered on mission, geography, or command.

Self-elected servers provide the same robustness for information dissemination that the peer-to-peer network transport layer brings to the transport layer.

SAIM provides large volumes of geolocation data — too much information for a human controller to monitor, particularly in high-stress situations. The EA alleviates this problem by using a machine-understandable plan to filter the information from SAIM and alert the user when events threaten the user or the execution of the plan. A plan-aware, situation-aware, action-specific EA can alert appropriately for the situation, thus improving decision making by enabling hands-free operations, reducing the need for human monitoring, increasing the amount of relevant information monitored, and prompting the user when action is required.

The complexities of plans, the number of agents, and the volume of data pose a challenge to existing execution-monitoring techniques. Unlike a lot of AI planning work, particularly in robotics, most actions in our domain are performed by external agents, mostly humans, and the monitor has no access to the state of executing agents. Status information must be obtained from external inputs.

We focus on the problem of alerting human users when the situation requires attention; we assume that the human will modify the plan as needed. This was done for several reasons. First, the users are unwilling to cede decision making to a machine, so we first develop trust by giving useful alerts, a capability well suited for automation if the plan can be represented with enough fidelity, and something that provides obvious value in dealing with the information glut. Second, mistakes can be a matter of life and death, so systems must be verifiably robust before they are given decision-making power. Human decision makers must take imperfect information into account, including reports from sensor networks, other humans, and execution assistants. Third, demonstrating the utility of automated, plan-based monitoring in this large and complex domain is likely to facilitate future acceptance by users of plan-related automation.

| name | Abbrev | Entities controlled |
|---|---|---|
| Battalion | BN | 400-600 |
| Company | CO | about 100 |
| Platoon | PLT | about 30 |

Figure 2: Echelons in the command hierarchy with EAs.

Execution monitoring requires coordination over multiple echelons (levels in the hierarchy), so that users know what their subordinates are doing. Figure 2 shows the echelons for which we have demonstrated the EA. Multiple agents at each echelon must coordinate fast-paced activities over a wide area in real time. Our task requires the solution of three difficult problems: handling the large volume of incoming information, developing a sufficiently rich plan representation for capturing tactical Army plans, and determining when to alert the user.

As mentioned before, the EA must give only high-value alerts to be useful. For example, once a unit is out of position or late, the system must recognize both the import of this condition and when the situation has changed sufficiently to issue another alert, without issuing too many alerts. Consider the seemingly simple example of a plan specifying that a squad of 10 agents should move to Objective Golf at 0700. What is the location of the squad? An obvious solution is to compute the centroid of each member's location. However, no one is near the centroid if all members are in a large semicircle with the centroid at the center (this situation arises when the squad follows a road around a sweeping curve). If one member is now immobile with his GPS still broadcasting, the centroid may be seriously inaccurate. Does the centroid need to be near Golf, or is one member near Golf sufficient, or must all members be near Golf? It depends on the mission (task) and situation. If the mission is to observe a valley, one member is sufficient, but we might want all members for an attack. Our solution is to use mission-specific algorithms (specified in the mission model described in Section 6.5) for reasoning about the location of units.

The EA must avoid cascading alerts as events get progressively further away from expectations along any of a number of dimensions (such as time, space, and resource availability). In the above example, how close in time to 0700 should the squad be before there is a problem with achieving the plan's objectives? Similarly, how close in distance to Golf? Again, the time and distance thresholds that indicate a problem depend on the mission and situation. A human uses his background world knowledge to quickly determine if a delay affects the plan, but execution aids must have much knowledge encoded to do this. These problems become exacerbated as the plans and missions become more complex. Detecting friendly-fire (fratricide) risks poses even more difficult issues, because there are typically many friendly units in close proximity.

## 6.2 SUO Approach

Machine understanding of the plan is the key to helping humans deal with the information glut created by advanced situation-awareness systems like SAIM. The plan specifies expectations for how events will unfold, so the EA can compare actual events to the situations that were anticipated. We use rich, knowledge-based plan representations (Wilkins & desJardins, 2001) to allow computers to share context with users, so both understand the semantics of plans and requests.

We had two tasks involving significant knowledge acquisition and domain modeling: (1) we had to model SUO plans and the actions that compose them, and (2) we had to model the value of information and various types of alerts for users. We interacted with several domain experts to develop these models. These tasks were aided by the centuries of analysis and modeling that have already been done in this domain. For task 1, the Army already has a standard plan representation called the *Operations Order*, which has a required structure, but the entries are mostly free text. Primitive actions in this domain are referred to as *missions*, and there are Army field manuals that describe missions in detail. We modeled missions in a hierarchical *mission model*. Our mission model and plans are described in Section 6.5. For task 2, there is extensive accumulated experience and analysis of errors and opportunities that arise during execution of SUO plans, but there are many tradeoffs to be made. The tradeoffs and our models are described in Sections 6.4, 6.6, and 6.7.

Mission-specific execution monitoring is achieved by a novel integration of mission knowledge represented as methods with an AI reactive control system. The EA invokes methods at appropriate points during plan execution. The methods employ mission-specific algorithms and in turn invoke EA capabilities in a mission-specific manner. Much of the domain and mission knowledge is encoded in the mission model and not explicitly represented in the plan itself, which specifies a partial order of missions for each team member. The EA uses the plan to invoke the knowledge in the mission model at the appropriate time and with the appropriate arguments.

Another feature of our approach, particularly for terrain reasoning, is the pervasive use of specialized programs, possibly external to the EA, to perform complex computations that are important to system performance. By using alternative specialized programs, the EA can easily adapt the granularity of its reasoning and improve performance as better modules become available. For example,

API functions in our design can be used for terrain reasoning and to compute the enemy strength from the current tracks.

Our approach builds on SRI's continuous planning technology (Wilkins & desJardins, 2001; Wilkins & Myers, 1998; Wilkins et al., 1995) and on the domain-independent Act formalism (Wilkins & Myers, 1995). Act represents procedural knowledge and plans as *Acts*, provides a rich set of goal modalities for encoding activity (see Section 6.5), and has been used by several institutions (Wilkins & Myers, 1998; Durfee et al., 1997). The EA uses PRS (Georgeff & Ingrand, 1989; Wilkins et al., 1995) as its reactive control system (other reactive control systems have similar capabilities, e.g., UM-PRS (Durfee et al., 1997)). PRS is a good framework for constructing the EA because it supports parallel activities within an agent, and can smoothly interleave responses to external requests and events with internal goal-driven activities with its uniform processing of goal- and event-directed behavior. PRS uses procedures encoded as Acts and its extensive graphical tracing provides valuable insights into EA operation.

### 6.3 SUO Architecture

The architecture of the EA and its interactions with the SAIM system are shown in Figure 3. We developed two major modules, the Planning Assistant (PA) and the Execution Assistant (EA), which assist the user in generating and executing plans, respectively. We implemented only a skeletal PA to produce machine-understandable plans, using the SIPE–2 hierarchical task network planner (Wilkins et al., 1995). Both the PA and EA use Acts and a common knowledge base, ontology, and mission model that is object-based and easily extended. Knowledge about actions is represented in the mission model, and knowledge of plans, strategies, and procedures is represented as Acts.

The inputs to the EA are plans to execute, location reports, sensor tracks, and messages from other agents (e.g., reporting mission success and failure, and ordering execution of new plans). SAIM broadcasts up-to-date locations of all friendly agents, and broadcasts tracks that represent the results of fusing sensor hits on nonfriendly entities. SAIM provides and the EA supports rates of more than a dozen such inputs per second.

The EA monitors the current mission for every immediate subordinate of the EA owner, and alerts on threats to subordinates (subordinate depth is customizable). If events threaten successful execution of the plan, threaten the user or subordinate units, or trigger planned contingencies, the EA issues an alert to the user, depending on the value of such an alert as determined by applying our VOA algorithms. The user must decide how to respond. Our design and technology can also suggest responses and/or plan modifications (Wilkins et al., 1995), but this was left for future work. In addition to giving alerts, the SUO EA can dynamically change the command hierarchy, abort execution of one plan and switch to monitoring a new plan, and reduce unwanted alerts to avoid inundating the user.

EAs for every unit at every echelon process reports and give alerts locally. SAIM provides the same tactical picture to all EAs (modulo an EA's registration in SAIM multicast groups). Therefore, it is not necessary for an EA to report a new threat to its superior, as the superior's EA (as well as the EA of other affected team members) has the same information and would already have issued a similar alert. This architecture is fault tolerant because EAs do not rely on reports from subordinates to determine most alerts. Thus, each EA maintains most of its functionality even if it is not in contact with other EAs, as long as it gets SAIM position reports from one node.

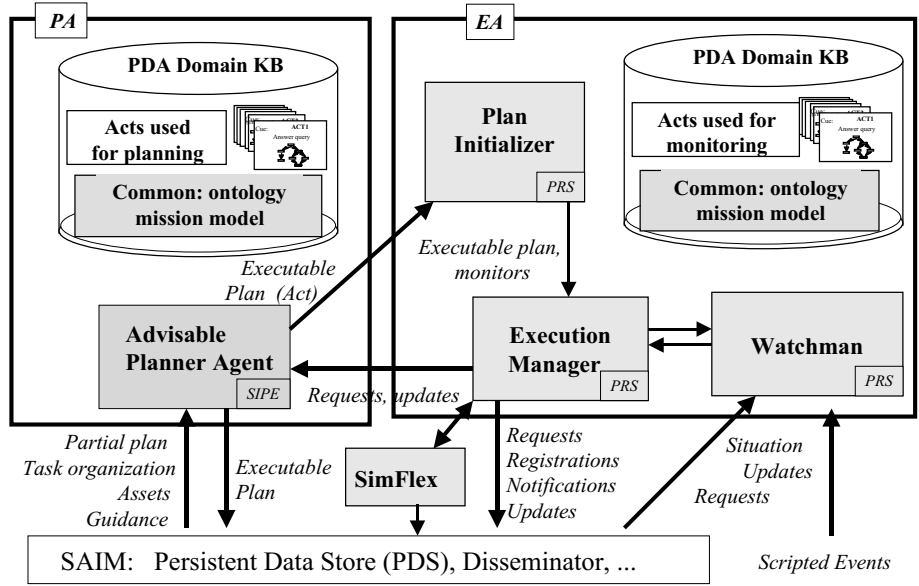

Figure 3: Internal architecture of the EA and PA and their interaction with SAIM. The PDS archives plans and other data and has a continuously changing picture of the current situation.

As shown in Figure 3, the EA is implemented as multiple asynchronous PRS agents (defined below) to alleviate the computational burden on the central EA Manager agent. Asynchronous agents provide faster response and better alerts than would a synchronous architecture, because the agents are always using the latest information available to them without having to wait to synchronize with other agents. To implement the EA, we extended PRS to monitor temporal constraints and to batch incoming facts so it could handle much higher data rates.

Internal EA agents (as opposed to external team members) use the Belief-Desire-Intention (BDI) model of agency (Rao & Georgeff, 1995). Each agent has beliefs about the state of the world, desires to be achieved, and intentions representing actions that the agent has adopted to achieve its desires. Each EA agent has its own controller process, which operates on its own database of beliefs, its own set of intentions, its own monitors, its own set of Acts that encode procedural knowledge about how to accomplish goals, and its own LISP functions that implement the primitive actions of the agent. An EA agent continually applies its Acts to accomplish its current intentions (tasks). The EA appears as a single agent to SAIM and the outside world. The following are the internal EA agents.

**Plan Initializer.** This agent gets a plan from the PA and sends messages to the EA Manager agent after performing all initializations necessary to begin monitoring of plan execution. Primarily, this involves creating and loading plan monitors, and posting facts in the EA Manager database.

**Watchman.** This agent monitors incoming message traffic on the SAIM network, mainly by querying for tracks and other information. It filters irrelevant or insignificantly changed reports, and sends a message to the EA Manager when any report or message requires its attention. It simultaneously monitors files of scripted events when such monitoring is requested. The Watchman inserts events from scripts at appropriate times, interleaving them with live messages.

**EA Manager.** This agent begins plan execution immediately after receiving a plan from the Plan Initializer. The agent implements the core EA functionality — it compares reports from the Watchman agent to the plan and plan monitors, and generates high-value alerts.

**SimFlex.** This agent provides a powerful and flexible way to define the execution semantics of an action using Acts (and thus the full power of PRS). SimFlex (simulated, flexible execution) enables mission-specific execution monitoring (by having an Act for each mission) and makes the system easily extendible. For example, if certain missions were to automatically command robotic vehicles or send messages to other EAs, those actions could be easily implemented in SimFlex. Most of the actions in our plans are executed by external human agents, in which case this agent does little except perhaps prompt the user. Some actions, such as reorganizations, are automatically executed in SimFlex by invoking the Execute-Mission method defined in the mission model.

## 6.4 SUO Alert Types

While there is extensive accumulated experience regarding the execution of SUO plans, selecting which types of alerts to detect involved trading off several factors, such as whether the alert can be detected from available data, the utility of the alert to the user, the cost of implementation, and the ability to maintain reactivity given the computational expense of detecting the alert. We earlier gave the example of balancing the usefulness of fine-grained terrain reasoning for movement projection with its computational impact on reactivity. Thus, modeling the value of information and types of alerts to be detected involved interaction between the domain experts and system developers. Here we describe the types of alerts we decided to detect. More details of how we implement these and model the value of information and alerts is given in Sections 6.6 and 6.7.

Figure 4 describes the 13 types of alerts that are detected by the SUO EA. Most of these are time and location checking. Comparing these alerts to our categories in Section 4, the proximity alerts are all instances of *adversarial activity detected*. The adversarial alerts also fit this category and the last three adversarial alerts are also of the type *plan constraint violated* because expectations and requirements specified in the plan (such as locations and routes to monitor) are violated. The contingency alert, which can be triggered by either friendly or hostile actions, is of the type *contingency plan suggested*. The out-of-position, coordination, and schedule alerts are of type *plan constraint violated*, but would be of type *constraint violation projected* when the violation is projected. The fratricide alert is of type *policy constraint violated*, and the unknown-position alert is of type *reporting requirement*.

## 6.5 SUO Plans and Mission Model

Our hierarchical *mission model* specifies an ontology for the primitive actions, and has methods that encode most of the domain knowledge about constraints and expected behaviors. Tailoring monitoring to each mission is crucial because most behaviors, even something as simple as denoting the location of a unit, are mission specific. The plan representation is a novel combination of the mission model and an extended version of the Act formalism (Wilkins & Myers, 1995).[1]

---

1. EA plans represent plans as they are expressed in Army operations orders, but only parts of the current Army five-paragraph order are represented in machine-understandable form. Primarily, task organization and the specific maneuver tasks and coordinating instructions from the Execution Paragraph are represented, but some other aspects are encoded as well.

| Alert Type | **FRIENDLY ALERTS** |
|---|---|
| fratricide | Friendly units pose a threat to each other. |
| out-of-position | Location constraints in plan violated. |
| unknown-position | Unknown location of a subordinate/coordinating unit. |
| coordination | Coordinating units cannot synchronize as planned. |
| schedule | Time constraint in plan violated. |
| contingency | An event has triggered a queued contingency. |
|  | **ADVERSARIAL ALERTS** |
| contingency | An event has triggered a queued contingency. |
| monitored | Activity at a monitored map location. |
| ave-of-approach | Activity on a monitored route (avenue of approach). |
| hostile-expected | Expected hostile activity absent. |
|  | **PROXIMITY ALERTS** |
| contact | A friendly unit's first contact with a hostile entity. |
| distance | Hostile entities are closer since last alert. |
| strength | Threat has grown stronger since last alert. |
| proximity | A merged alert of more than one of the above. |

Figure 4: Types of alerts generated by the SUO EA.

The Act formalism is a domain-independent AI language for representing the kinds of knowledge about activity used by both plan generation and reactive execution systems. It provides a rich set of goal modalities for encoding activity, including notions of achievement, maintenance, testing, conclusion, and waiting. This expressiveness is necessary for representing SUO plans, which must coordinate distributed units, trigger preplanned contingencies, and enforce a variety of execution constraints. The basic unit of representation is an *Act*, which can be used to encode plans, strategies, and standard operating procedures (SOPs).

The EA can monitor any plan that is composed of missions from the mission model. The mission model is derived from Army field manuals and elaboration by domain experts. It includes a set of mission templates (with associated parameters) that units at various echelons could be ordered to perform, in either a written or verbal order. Since the mission model is grounded in field manuals, it is a first step toward formalizing a plan representation that is meaningful to end users yet amenable for execution monitoring and other AI-related capabilities (e.g., plan generation, replanning, course of action evaluation).

The mission model is a class hierarchy (implemented in LISP and CLOS, the Common Lisp Object System), with inherited methods that encode knowledge about how to monitor a particular mission. Each leaf class corresponds to a monitorable action that may occur in a plan; each nonleaf class encapsulates common parameters and behaviors for its subclasses. The mission model allows most aspects of system behavior to be tailored in a mission-specific manner. Thus, specialized methods in the mission model can, for example, use mission-specific algorithms for monitoring progress of a movement. Methods are invoked by the EA Manager but can in turn invoke processing in the EA Manager by posting mission-specific facts that invoke capabilities of the EA Manager (there is an API of such facts, important facts are described later).

Each mission in the model contains a name and parameters that describe the mission. For example, the mission model contains the nonleaf **movement-mission** class, which contains a destination parameter and a method for checking that the executing unit has arrived at its destination. Five different movement missions inherit this behavior. The root class in the model is the **mission** class, which encapsulates all parameters and behaviors that are shared by all missions. All missions inherit start-time and end-time scheduling constraints and methods from this superclass.

**Coverage.** The mission model formalizes a substantial subset of the missions mentioned in Army field manuals. We have enumerated 62 mission classes, and have implemented 37 of these, a superset of those required by our scenarios. The mission model covers multiple echelons, with emphasis on battalion, company, and platoon. It does not model all aspects of missions, only those for which SAIM can provide monitoring data, that is, those related to time and location. For example, it does not alert on potential mission failure due to casualties incurred.

**Contingencies.** The mission model contains the nonleaf **contingent-mission** class. This class and its leaf children classes are used to implement a mission sequence that is part of the plan but is to be executed only when certain conditions are fulfilled. Domain experts term these portions of the plan *branches* and *sequels*. The missions under **contingent-mission** contain parameters to describe the condition, specified in the plan, that activates the contingency.

**Dynamic resubordination.** Army operations orders allow the command hierarchy (termed the *task organization*) to be changed during the operation, although existing command and control software does not support dynamic changes to the command hierarchy. The **reorganization-mission** class provides this capability in the EA. When a reorganization mission is executed, it causes the EA to update its representation of the command hierarchy accordingly. This has a substantial effect on EA behavior, because many EA algorithms use the command hierarchy.

**Methods.** Each mission provides several methods that are invoked at appropriate times by the EA to monitor execution of the mission. This set of methods serves as an API for mission-specific execution monitoring semantics. The following methods comprise the bulk of the API:

**Post-Execution-Constraints** is the main API method invoked by the EA for monitoring a mission. It invokes methods that post and enforce various constraints.

**Check-Initial-Location, Check-Final-Location** confirm that unit(s) are positioned correctly at the start and end of their mission respectively.

**Start-Time-Constraints, End-Time-Constraints** check that a mission is beginning and ending execution as scheduled. These methods usually post facts in the EA Manager to invoke its Timed Monitor mechanisms.

**Location-Constraints** enforces location checking of friendly units and hostile tracks for a variety of missions.

**Contingency-Satisfied** determines whether a contingent mission sequence should be executed.

**Respond-To-Monitored-Red-Activity** is the algorithm for responding to hostile activity in places where the plan calls for monitoring such activity.

**Execute-Mission** invokes any processing required to execute a mission. It is invoked by posting a goal in the SimFlex agent, and all internal agents continue PRS execution while Execute-Mission is running.

**Compute-Priority** computes the priority of an alert.

**Desired-Strength-Ratio** is a heuristic that expresses a desired friendly:hostile ratio of combat power.

**Red-Alert-Priority** computes the priority of a proximity alert, or whether an alert should be issued at all, based on recent changes to reported strengths of a friendly unit and nearby hostile tracks.

**Wait-Until-Mission-Start, Wait-Until-Mission-End** control interaction with the EA GUI with regard to mission start and end times.

Specialization of methods is useful for expressing desired behavior by the EA. For example, the Location-Constraints method is specialized on **movement-mission**, **coordination-mission**, and several other missions. For movement missions, the EA checks whether the centroid of the moving unit is at its destination. For coordination missions, the EA checks whether any elements of two coordinating units are at the specified coordination point.

### 6.6 SUO Execution Monitoring

The EA Manager continuously responds to new goals and facts posted in its database. The Watchman agent is asynchronously posting facts to the EA Manager database as it receives messages from SAIM. Facts so posted include confirmations of mission starts and completions (from subordinate EAs), orders for aborting the current plan or executing a new plan (from a superior EA), sensor tracks, calls for fire, and location reports (from the SAIM network).

The methods in the mission model post facts to the EA Manager to invoke mission-specific monitoring. Examples of such fact-invoked capabilities provided by the EA Manager include monitoring several types of time constraints and monitoring a specified location for activity (with options for friendly or enemy, and expected or unexpected).

The behavior of the EA Manager is determined by the posted goals and facts, their relative timing, and the set of Acts used to respond. The EA Manager switches its focus to the highest-priority task on each execution cycle so that all goals and facts generate responses with acceptable latency (Georgeff & Ingrand, 1989). Execution cycles are on the order of milliseconds. System behavior is nondeterministic because it depends on exactly which facts and goals are posted during each execution cycle, which may in turn depend on the CPU scheduling of the EA Manager, Watchman, and SAIM processes. The number of alerts rarely varies — what does vary is the exact times of alerts (which can vary by a few seconds), and the hostile strengths reported (which can change if the Watchman agent gets more or fewer CPU cycles to accumulate tracks before the EA Manager executes).

The EA Manager must constantly monitor the status and behavior of currently executing missions, while simultaneously monitoring up to a dozen incoming facts per second and determining their impact on the plan. While monitoring a plan, it typically has on the order of 100 intentions it is trying to accomplish at any one time, and has 107 Acts (Procedures) to apply to its intentions. Most intentions can cause an alert to be generated. Each unprocessed report and track forms an intention. Typically, five subordinate missions are executing simultaneously. Each produces multiple intentions: at least one for detecting the start and end of each mission, and a few for each time and location constraint (every mission has at least a start time and an end time constraint). For example, for each time constraint, the EA Manager has intentions that monitor if the specified time has elapsed and if the required event has occurred.

The plan-based monitoring of the EA can be viewed as asynchronously and simultaneously interleaving the following activities. We describe these in more detail below and mention the most important design tradeoffs.

- Initiating or aborting plan execution upon request

- Monitoring incoming location and sensor reports

- Monitoring progress of the missions and time constraints specified in executing plans

- Responding to other types of incoming requests

### 6.6.1 PLAN MONITORING

To monitor a plan, a request goal is posted in the database. This invokes an Initialize-Plan Act that computes the conditions that should be globally monitored for this plan and posts facts to the EA Manager database declaring that there is a current plan with monitors. These facts in turn cause Acts to execute in the EA Manager, which load and execute the plan. The EA Manager traverses through the parallel branches of this plan as missions complete.

The global monitors are computed using the API function *compute-plan-monitoring-data*, which can specify domain-specific monitors. Domain-independent capabilities are also available, such as having the system determine all predicates in plan preconditions that must be true initially (as opposed to predicates that are achieved by plan actions that precede them). In the SUO domain, *compute-plan-monitoring-data* finds all decision points and "named areas of interest" specified in the plan, and sets up monitors for them. This monitoring is accomplished by posting facts in the EA Manager database that cause the EA to notice any adversarial activity in these locations.

The EA can abort monitoring of one plan and switch to monitoring a new plan. This process involves removing facts for old missions and monitors from the EA Manager database, aborting execution of the Acts currently intended for execution, and posting a goal to execute the new plan.

### 6.6.2 LOCATION REPORTS

The Blue Report Act in Figure 5 is invoked every time a location report is posted in the EA Manager database, which can happen several times each second. However, the Watchman agent filters location reports that are not of interest to the EA Manager (e.g., for entities irrelevant to the plan of the EA owner, or because there is no change from the last report), and updates the representation of the current situation in the EA. The Blue Report Act is specific to the SUO domain, but our framework requires a similar Act to be written for each type of input that is to be actively monitored. For example, there is a similar UV-Robotics Act that responds to state updates (see Section 7.4). These Acts are written using the Act-Editor (Wilkins & Myers, 1995), a tool for graphically editing procedural knowledge (Acts) with an intuitive user interface.

This Act begins by invoking a domain-specific specialized reasoner to check for fratricide risk, which may have the side effect of giving an alert (using the API function *issue-alert*). The specialized reasoner can easily be replaced by better fratricide detection algorithms in the future. Next, the Blue Report Act checks whether the current plan has any expectations for this unit, and if so, it calls the API function *check-expected-location* to compare the current location to the expected location, again posting an alert if appropriate. Finally, the report fact is removed from the database.

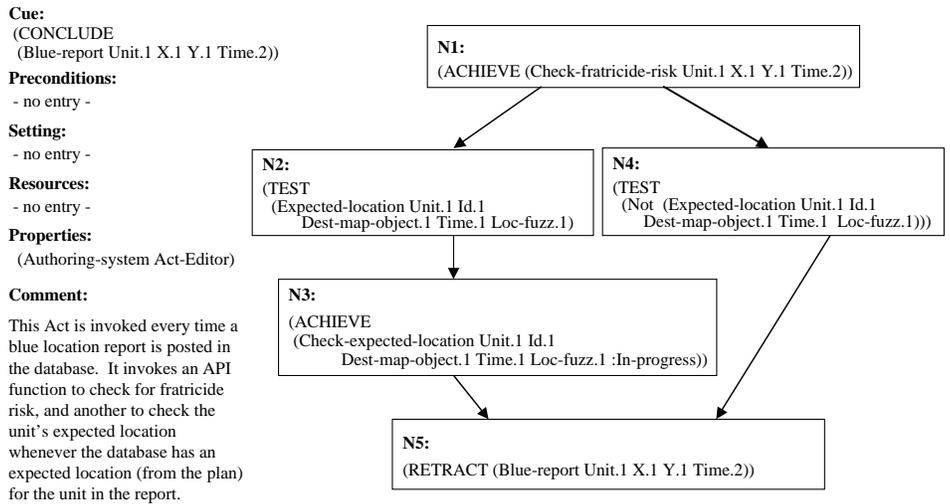

Figure 5: Graphical representation of the Act that responds to every friendly location report.

Responding to a fused sensor track indicating adversarial activity is controlled by a similar Red Report Act, which compares adversarial activity to plan expectations. Instead of analyzing fratricide risk, the Red Report Act invokes a reasoner for evaluating adversarial threats. As described in Section 6.7, this involves updating a *threat envelope* for each friendly unit.

### 6.6.3 MISSION MONITORING

To explain mission monitoring, we give an example of how a move mission in a plan is monitored. A **move-mission** ready for execution has the following parameters:

(**move-mission** *unit start-time-constraint start-time end-time-constraint end-time destination route formation march-technique contingency contingency-satisfied*).

The EA Manager begins execution by calling three methods defined in the mission model: Start-Time-Constraints, End-Time-Constraints, and Location-Constraints. Each of these posts facts in the EA Manager database to invoke mission-specific monitoring capabilities. For example, Location-Constraints (which is specialized to the class **movement-mission**) posts facts about locations this mission expects friendly units to occupy and at what time (derived from the *destination* and *route* arguments), and might also post facts about locations where this mission expects adversarial activity and where adversarial activity should be monitored/alerted.

The EA receives confirmation of mission start from a subordinate EA. Location reports are continuously posted by the Watchman, and the Act in Figure 5 analyzes them with respect to the location facts posted by Location-Constraints. Sensor tracks are similarly analyzed by a different Act. Let us suppose that at some point during mission execution, a track shows activity in a location monitored by this mission. The EA would detect this and invoke the mission-specific method

Respond-To-Monitored-Red-Activity, which describes how this mission will respond to such an event. For example, it could issue an alert, abort the move, execute a contingency plan, or ask the user to choose from a set of such options.

| Type | Meaning |
|------|---------|
| ASAP | start/end not specified |
| ON-ORDER | start/end when ordered |
| START-AT | start exactly at given time |
| START-NLT | start no later than given time |
| START-NET | start no earlier than given time |
| END-AT | end exactly at given time |
| END-NLT | end no later than given time |
| END-NET | end no earlier than given time |

Figure 6: Temporal constraint types.

### 6.6.4 TEMPORAL MONITORING

The mission model includes starting and ending time constraints for every mission. Each time constraint consists of a *temporal constraint type* and an absolute time. The temporal constraint types in the EA are shown in Figure 6. These constraints require two types of monitoring tasks: detecting when time constraints in the plan have passed without being met, and detecting events that occur before their specified time.

We extended PRS with a domain-independent *Timed Monitor* mechanism that provides a general capability covering all our temporal monitoring requirements. This capability was implemented in the form of Acts, with some supporting LISP code. Four special types of timed monitors are provided, invoked by posting facts with the predicates Check-Not-Later-Than, Check-Not-Earlier-Than, Check-In-Window, and Check-Near-Time. We describe our implementation for one of these; the others are similar. The Act Check-Near-Time checks that an event occurs within a specified threshold of some time point and can be invoked by a fact of the form:

*(Check-Near-Time event.1 time.1 mode.1 fuzz.1)*

To succeed, *event.1* must occur within *fuzz.1* seconds of *time.1*, with *mode.1* indicating whether this time is absolute or relative (to the time at which this fact is posted). A Timed Monitor Act sets up a timer that expires at the given time, and PRS reacts appropriately to either the expiration of the timer or the occurrence of the event, posting facts to the database to note the success or failure of the temporal constraint. Because the above Acts are fact invoked, these mechanisms enable the establishment of separate intentions to perform timing, without blocking other processing. This modularization enables triggers to be set up to independently respond to timing results.

### 6.6.5 DESIGN TRADEOFFS

As described in Section 2, a balance must be struck between the capabilities provided and resources used. The tradeoffs are different in every application and are usually a critical aspect of the design of an execution assistant. In the SUO domain, terrain reasoning is a key factor in this tradeoff. Using fine-grained terrain data to analyze progress or project future failures can overload computational resources. Therefore, the EA uses coarse terrain reasoning, but our design allows higher-fidelity terrain reasoners to respond to a defined set of terrain analysis requests. This feature allows the system to adjust its analysis to the tempo of operations.

Other key features to consider when making tradeoffs between reactivity and capabilities are the amount of processing done by the mission-monitoring methods, the report-monitoring Acts, and any specialized reasoners (such as terrain reasonsers) invoked by the methods or Acts. The user can adjust the frequency of monitoring at any time by customizing parameter settings. Currently, the SUO EA is not computationally overburdened while analyzing every report in full, but adding more computationally expensive projections or alerts in the future could cause reconsideration of this design decision. Finally, the amount of filtering of incoming reports done by the Watchman agent affects this balance.

### 6.6.6 OTHER FEATURES AND IMPLEMENTATION

The EA responds to other requests, such as calls for fire, which are described in Section 6.7. Several other capabilities were implemented to make the EA easier to use and understand. Two are briefly mentioned here. We implemented a GUI, not meant for military users, but rather to facilitate evaluation and understanding of the EA. The GUI displays all alerts in different scrollable windows for each priority level, the current time, and the current mission of each subordinate of the EA owner.

The user can confirm mission starts and ends locally, although this might be done with voice or some other modality in a fielded system. When a confirmation arrives from a subordinate EA, the confirmation window for that mission is destroyed. Thus, confirmations and prompts can be given locally or received in messages, with a seamless interleaving of those two types of confirmation.

The EA, PA, mission model, PRS and SIPE–2 are implemented in COMMON LISP, CLIM, and CLOS. The EA also contains procedural knowledge in the form of Acts. SAIM was implemented in C++ and Java, using the ACE Object Request Broker for CORBA. C++ was used to interface the EA to SAIM and CORBA.

## 6.7 Alert Detection and VOI/VOA

The central task of the EA is to notify the user of important changes to the situation that may demand attention. The EA must also avoid excessive alerting; otherwise, the user would abandon the EA as a nuisance. A model of the user's cognitive state with respect to awareness of threats would be ideal, but is unavailable. As described in Section 5, we developed algorithms that heuristically estimate VOI and VOA using domain knowledge. The inputs to these algorithms are described in Section 5.3. We avoid excessive alerts by issuing only high-VOA alerts. Our techniques include

- Keeping event histories for each friendly unit, for map coordinates and for important map locations named in the plan (e.g., decision points).

- Having alerts "expire" in the sense that they can no longer be used to suppress future alerts.

- Using alert histories for suppressing alerts by time (similar alert given recently), strength (threat not significantly stronger), and distance (threat not significantly closer).

- Merging several related alerts that apply to subordinates into one alert for the common parent.

- Providing parameters so the user can customize alerting behavior and VOI/VOA estimates.

The event histories are currently our only model of the user's cognitive state, except for global properties of the situation, such as operational tempo. Our VOA calculations take into account the frequency and timing of alerts that have already been given. The histories include all alerts that were issued to the object of the history, and may include additional events, as described in Section 7.4. We assume that the user is aware of information about which he or she has recently been alerted. The idea behind having alerts "expire" is that the user may have forgotten information provided too far in the past. Thus, the EA will not use alerts older than a specified threshold to reduce its estimate of the value of giving an alert now.

The EA's behavior must be easily customizable, both by users and by the plan, because users have different preferences and situations impose different requirements. The EA can be customized in many different ways. Our VOA algorithms, which recommend alerts and classify them by priority level, are controlled by thresholds and repetition parameters, which allow alerting behavior to be customized to the user or situation. Examples of customizable VOA parameters are the alert expiration periods described above (default 12 minutes) and alert suppression intervals (90 seconds for hostile alerts, 120 seconds for alerts about friendly team members) during which alerts of the same type about the same objects are suppressed for the given interval. In terms of VOA, another fratricide alert has no value for the first 120 seconds after the user has been alerted about a fratricide risk from the same team member.

Examples of customizable VOI parameters are the out-of-position distance threshold (150 m), thresholds on the strength of adversarial threats, and the time threshold for schedule alerts (30 seconds). The time threshold, for example, would be smaller for tightly coordinated operations, and larger for more loosely coordinated plans. In terms of VOI, detecting that a team member is late has no significant value until the tardiness reaches the given 30-second threshold. If certain missions in the plan change this threshold, say to 10 seconds, it indicates that information about tardiness of 10 to 30 seconds has more value in the context of these missions.

The problem of avoiding unnecessary repetition of similar alerts occurs with every type of alert. Schedule deviations can become progressively more off schedule, position deviations can become progressively more out of position, threats can move progressively closer or become progressively stronger, and fratricide threats can persist over time. An EA must avoid cascading alerts in each of these cases. In our framework, customizable thresholds are often paired with either customizable ratios or a customizable sequence of thresholds, which control how often to repeat the alert if the mission deviates progressively more from expectations. Repeated alerts generally have a lower VOA and are given lower priorities.

Our evaluation showed that two types of alerts in the SUO domain pose particular problems for avoiding inundation of the user. These are proximity alerts about adversarial activity and alerts about fratricide risks among team members. We developed VOI/VOA algorithms especially for these two types of alerts.

### 6.7.1 PROXIMITY ALERTS

There can be a high volume of sensor tracks near friendly units prior to and during battle; it would overwhelm the user to see an alert on every change to every track. We keep a *threat envelope* for each friendly unit, consisting of tracks close enough to pose a threat to it. Tracks are placed in zero or more threat envelopes when they appear or move. Only significant changes to the strength of the aggregate force in an envelope or the closeness of the nearest track causes an alert.

### 6.7.2 FRATRICIDE RISKS

Fratricide is one of the biggest dangers on the modern battlefield. This risk increases as the range, lethality, and accuracy of weapons increase. Increased range increases risk because there is a bigger area in which every team member must be correctly identified. Increased accuracy increases risk because an incorrectly targeted team member is more likely to suffer harm. Hopefully, tools like the EA and SAIM will increase situational awareness and greatly reduce the frequency of incorrect targeting. Usually, a large number of friendly entities are in close proximity, so many potential fratricide situations exist.

The EA detects two types of fratricide risks: (1) from calls for fire from other team members (which appear in messages from SAIM), and (2) friendly units near each other (which are detected from the geolocation data). In the first case, the user who issues a call for fire is warned and asked for confirmation if team members are within a given threshold of the target. If the request is confirmed, a SAIM message is sent to team members, and the EA of any entity within the target threshold immediately alerts its owner to the risk from the planned fire.

The second case produces far too many alerts if simple algorithms are used. Our algorithms are based on the Army's notion of unit boundaries, which are specified in the plan. When two units are within their boundaries, no alert is issued even if they are within weapons range of each other. Fratricide alerts are issued when one unit is in another unit's boundaries and within weapons range of the other unit. We handle numerous special cases, such as when two units are both outside their boundaries and within weapons range of each other. Detection of other fratricide situations is left for future work (e.g., misoriented units within their boundary).

## 6.8 SUO Evaluation

The EA was evaluated with respect to the usefulness of its output, frequency of unwanted alerts, and real-time performance with realistic data streams. SAIM and the EA were tested against data produced by a high-fidelity military simulator on two scenarios. The simulator has detailed models of each type of vehicle and sensor. One scenario lasted 13.5 hours, but only the last 90 minutes were simulated at high fidelity. (The first 12 hours had a file of scripted events, with a few dozen tracks and reports.) The second scenario, on the same terrain, was simulated for 20 minutes. The 90-minute simulation had more than 45,000 events passed to the Watchman from SAIM, of which 13,000 were passed on to the EA Manager, which monitors only to the squad level (8 to 10 entities). During the simulator run, scripted events also simulate messages from any team members that are not running live (such as messages confirming mission starts and completions). The high fidelity of the simulation provides realistic data rates and inputs, thus providing some evidence indicating that the EA will perform as desired in the real world.

Our formal evaluation ran live with SAIM, the simulator, and several team members, each running their own copy of SAIM and the EA on different physical machines. For a shorter development cycle, we implemented an event generator that reproduces SAIM behavior, making the SAIM network unnecessary. The event generator creates messages from files of scripted events that include confirmations of mission starts and completions (that normally would come from a subordinate EA), orders for aborting the current plan or executing a new plan (that normally would come from a superior EA), and sensor tracks and location reports (that normally would come from the SAIM network). Our event scripts contain all messages captured from a run that included the simulator and SAIM.

### 6.8.1 QUALITY OF ALERTS

Figure 7 presents the total number of alerts by type at each echelon during a typical run. Flash is highest of four priorities, and "immediate" the second highest. Flash alerts are generally life threatening (first contact with adversarial entities and fratricide), while lower-priority alerts are only plan threatening.

We analyzed and evaluated the alerts generated from our first and most challenging scenario. Analysis by SRI and our domain experts indicates that all important situations were alerted. Less than 10% of alerts were judged to have such low value that they should not have been issued, and no Flash alerts were so judged. Judging the VOA for each alert is subjective: different domain experts may have different alerting preferences, and each alert will have some new information. We have no firm data on the number of unwanted alerts that would lead to performance degradation for a typical user (or that would cause a user to shut off his EA). It is clear that the 86% false-alarm rate found in a pediatric ICU (Tsien & Fackler, 1997) would not be acceptable on the battlefield. In our judgment and that of our domain experts, the rates of low-value alerts we achieved are acceptable. The number of alerts in Figure 7 is reasonable for a 90-minute interval of fast-paced action, and further elimination of alerts risks missing a high-value alert. We purposefully erred on the side of not missing any alerts.

We have compared the alerts generated by EAs operating at different echelons (running on different machines on the SAIM network) on the same simulation. Our analysis shows that they detect the same threat at the same time from the same tracks, when the threat is relevant to their plans. The alerts show plan-specific and mission-specific behavior as expected. Because of the nondeterminism inherent in our asynchronous agents, the alerts do not always show the exact same strength, bearing or location of a threat. Figure 8 shows one example, the BN and A CO alerts near 08:05. At this time, 2nd PLT, A CO is moving outside its unit boundary specified in the plan, and a hostile force appears to the north of A CO moving south. Note that both EAs issue flash fratricide alerts at 8:05. However, the other alerts are different, and specific to the plan and owner of that EA, as we would expect.

The plan called for an Attack-By-Fire mission if tracks are observed at location DP2 (a decision point at which hostile activity calls for a human decision). This immediate alert appears only on the BN EA (because the contingent fire mission is in only the BN plan) and notifies the user that hostile entities have entered DP2, triggering the contingency. AA-Diamond is a route, defined in the BN plan, along which adversaries are likely to approach. The second alert notifies the BN user (only) of activity on the route and reports the number of entities detected.

Both EAs independently identify the fratricide risk at 8:05, as would the EAs of the two platoons involved. The message details the two platoons to facilitate a quick response. Next, the BN EA issues a distance alert after detecting tracks 450m SE of the Recon PLT, which was subordinated to the BN earlier in the plan (so only the BN EA alerts). These tracks are now closer than when an earlier first-contact alert was issued. Finally, the out-of-position alert at 07:58 indicates that 2 PLT is 1 km south of the route specified for its move mission. (The 2 PLT EA simultaneously alerts that one or more of its subordinate squads are out of position.)

| Number | Type of Alert |
|--------|---------------|
| | **Battalion EA - 41 missions** |
| 78 | total alerts over 13.5 hrs, 26 flash |
| 33 | proximity alerts |
| 3 | schedule alerts |
| 2 | out of position alerts |
| 17 | avenue of approach alerts |
| 5 | triggers contingency alerts |
| 5 | at-monitored alerts |
| 9 | fratricide alerts |
| | **A CO EA - 11 missions** |
| 36 | total alerts over 1.5 hrs, 14 flash |
| 19 | proximity alerts |
| 3 | schedule alerts |
| 1 | out of position alert |
| 6 | avenue of approach alerts |
| 0 | triggers contingency alerts |
| 3 | at-monitored alerts |
| 4 | fratricide alerts |
| | **3 PLT, A CO EA - 6 missions** |
| 7 | total alerts over 1.5 hrs, 2 flash |
| 6 | proximity alerts |
| 1 | schedule alert |
| 0 | all other alerts |

Figure 7: Number of alerts by type at each echelon. The number of missions for each echelon indicates the size of its plan. Only the last 90 minutes of the 13.5-hour scenario was simulated at a high fidelity. Of the 78 Battalion alerts, all but 5 were issued over the last 90 minutes.

### 6.8.2 PERFORMANCE

Our EA Manager must handle more than 100 simultaneous intentions, while determining the import of a dozen or more new facts a second and checking alert histories for redundancy. It was not clear that our system could do all this and still alert the user within 5 seconds of a new fact arriving, as required by our users. We tested the EA on both scenarios to determine if it met these requirements. In real time, the EA generated alerts in less than 2 seconds from the receipt of a new fact. We found that the EA can not only keep up, but can run at between 10x and 20x real time. (There may be anomalous schedule alerts because of granularity issues at high time expansion rates.) Thus, current data rates are not close to stressing the system – at 10x real time we are processing an average of 24 events per second in our 90-minute simulation, which is double our design requirement of a dozen events per second. We did not determine the multiple at which degradation would occur because it is difficult to detect degradation in such a complex system. We did establish that the EA, using

**From BN EA** 0803-0805

DAY 2, 08:04 IMMEDIATE notification:

**Red activity at DP2 triggers contingency** for Attack-By-Fire

DAY 2, 08:05 ROUTINE notification:

**Enemy activity on ave of approach** AA-Diamond (8 vehicles)

DAY 2, 08:05 FLASH notification:

**Fratricide risk 2 Plt, A Co** moved out of position near 3 Plt, B

DAY 2, 08:05 IMMEDIATE notification:

**Closest threat (tracked) is now closer**- 450m SE of Recon PLT

**From A CO EA** 0758-0805

DAY 2, 07:58 IMMEDIATE notification:

**2 PLT out of position** for Move mission is at GL180837, should be at Line-0003 (1000 m. N of 2 PLT)

DAY 2, 08:05 FLASH notification:

**Fratricide risk - 2 PLT** is out of position near 3-3-B-2-66

Figure 8: BN and A CO alerts around 0805 on Day 2. There was only one A CO alert from 8:03 to 8:06, during which time there were several BN alerts.

100-meter map granularity, is easily sufficient for plan monitoring with SAIM data rates (running on both Sun Ultra 60s under Solaris and Pentium-based machines under Linux).[2]

Prior to implementation of the EA, we did a performance evaluation of PRS to determine if it could handle the input data rates required by the EA. We briefly describe our results as many other reactive control systems are based on PRS, e.g., UM-PRS (Durfee et al., 1997). We found that it could not handle more than 12 facts per second without unacceptably long delays, using randomly generated facts for two predicates where each fact invoked only trivial processing (incrementing a counter). We determined that the effects of combinatorial PRS algorithms could be avoided by batching new facts each time through its control loop. We modified the control loop to do so, and the performance improved remarkably. For a test case of 2,000 facts posted in 1 second, it reduced time to respond to the first (any) fact by 84%, reduced time to respond to all facts by 72%, and reduced memory usage by 83%. Experiments showed that a fact batch size near 55 was optimal for reducing response time, and any value between roughly 25 and 100 was near optimal.

### 6.8.3 LIMITATIONS

The EA is limited by what has been modeled, by the low fidelity of some models and heuristics, and by the scenario-specific population of the knowledge base. There are many aspects of plan execution that we do not currently monitor, although our monitoring framework can be easily extended when other aspects of plans are modeled. Our selected capabilities are mostly a function of the available input data and available funding for modeling. The EA can monitor a much broader range of plans than were used in our scenarios. In fact, it can monitor any plan composed of a partial order of defined missions for team members.

---

2. All performance data are from a Sun Ultra 60 under Solaris. All product and company names mentioned in this document are the trademarks of their respective holders.

# 7. Monitoring Robot Teams

We are using a team of robots to cooperatively track and pursue enemy entities that have been detected. Unmanned air vehicles (UAVs) and unmanned combat air vehicles (UCAVs) are a growing research interest (Musliner, Durfee, & Shin, 1993), led by the availability of cheaper platforms that are easier to use. The SRI UV-robotics project focuses on building a system to carry out a mission objective using a team of UGVs and UAVs. Each UGV or UAV is an autonomous agent with its own view of the world, own onboard reasoning capabilities, and own set of resources (such as power, computation, and a unique set of sensors). During a mission, there may be limited opportunity to communicate with the human controller. Therefore, the agents must rely on one another to complete the mission. Our research concentrates on providing reactive regulation of low-level sensor systems and vehicle controllers so as to attain high-level mission goals, while reacting to unforeseen circumstances and taking advantage of the evolving situation.

The UV-robotics domain resembles the SUO domain in that it requires the rapid assessment of the operational situation, the determination of the viability of existing plans and control policies, and the modification of goals and objectives based on those findings and the available resources. Unlike the SUO domain, the decisions are made by the (automated) agents themselves and the agents must negotiate solutions in a cooperative fashion. One of the challenges of UVs (or any physically mobile agent) is the need for a reactive system. Perception of, and knowledge about, events and actions in the physical world are generally imprecise. To perform tasks reliably and repeatedly requires dynamic monitoring.

Just as the SUO EA filters alerts to avoid overloading the human decision maker, we must also filter alerts to an autonomous agent to avoid overloading its computational resources. Resources are always limited, particularly on a mobile platform, so a balance must be struck between usefulness and resources used. A good example of such balance is the computational resources available onboard our robots. With an infinite number of CPU cycles, we would be able to generate large numbers of contingency plans and evaluate each with simulation. However, we have only 20% of the CPU available for robot control and monitoring. Therefore, we have to make design decisions that limit the complexity of both control and monitoring algorithms, possibly leaving extension hooks in anticipation of greater processing power in the future.

## 7.1 UV-Robotics: Problem Description

Our long-term goal is to build, test and validate an architecture for an agent that can support multiple goals in a dynamic environment of cooperative mobile agents. Initial tasks for our teams include surveillance and reconnaissance, search and destroy, pursuit, and evasion. A team of robots would be expected to perform these tasks with minimal supervision. Key components of this architecture were identified to be negotiation, strategic planning, execution and tasking control, execution monitoring, and recovery from failure. The challenge is to not only have several robots working together but to have them understand the effects of their actions on common team goals.

One challenge is that an agent may be working toward multiple, possibly conflicting, goals. Thus, the agent must be constantly evaluating its commitment to actions, or tasks, that contribute to the satisfaction of these goals. The imprecision of any action or sensory input has to be taken into account, and its contribution toward the satisfaction of current goals or plans assessed. In addition, the user must be kept informed of the progress of the team toward its goals. The user does not want to be actively involved in robot control, but must be able to intervene when necessary. Thus, monitoring must both ensure robust autonomous operation and provide the user with a window into the operation of the team.

## 7.2 UV-Robotics: Architecture

The SRI UV robot architecture is based on several years of research at SRI into intelligent reactive control, planning, negotiation, and robot motion control (Wilkins & Myers, 1995; Myers, 1996; Wilkins & Myers, 1998; Cheyer & Martin, 2001; Konolige & Myers, 1998). It is similar to systems like SAFER (Holness, Karuppiah, & Ravela, 2001) and SRTA (Vincent, Horling, Lesser, & Wagner, 2001) in its ability to deal with multiple goals at once and evaluate when to discard goals. Figure 9 shows our Multi-Level Agent Adaptation (MLAA) architecture. Clearly, monitoring is pervasive and serves each layer in the architecture as well as the user (not shown).

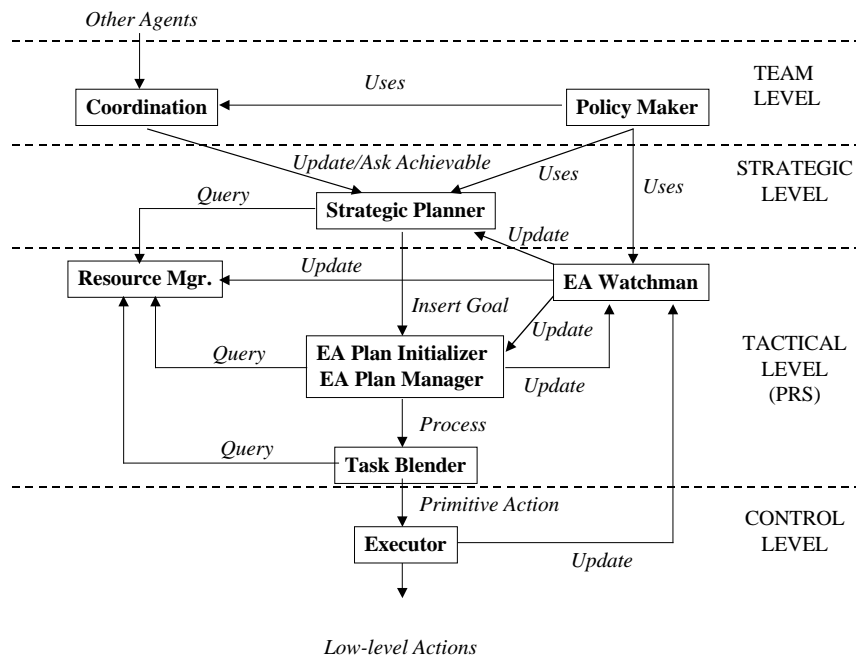

Figure 9: Multi-level Agent Adaptation Architecture.

The coordination module receives goal requests from the human commander or other agents. The agent participates in a negotiation process to determine its role in achieving the goal. During negotiation, the agent consults the strategic planner to create a plan, or plan segment (referred to as a *recipe*), and assess the recipe's viability given current commitments. If the negotiation process results in the goal and its recipe being accepted, the EA Manager (see Figure 3) instantiates the recipe and initiates its execution. The Plan Initializer also creates monitoring sentinels for use by the EA to detect deviation from the recipe during execution. The execution of a recipe involves activation of tasks that must be blended with other active tasks to maximize the satisfaction of multiple goals. For example, if the robot needs to reach a waypoint by a set time, take a picture of a location nearby, and also remain concealed, the task blender modifies the path planner at runtime to achieve all three tasks. Finally, the lowest layer in the architecture is the interface between the tasking architecture and the physical, or simulated, robot controller.

The monitoring in Figure 9 is done by the UV EA, which was created by using the architecture and representations of the SUO EA. The modular design of the SUO EA made this adaptation straightforward. The architecture and internal EA agents depicted in Figure 3 were used with little modification, as were the plan representation and the techniques for monitoring plans, applying VOI and VOA calculations, and issuing alerts. Our implementation of an initial UV EA (using code from the SUO EA) was done in about one person-week, an impressive result given the complexity of the task. The implementation included connecting to new data sources, parsing their messages, determining and implementing the most valuable monitoring algorithms, integrating with the plans and missions already defined, and writing domain-specific VOI/VOA algorithms. Achieving some missions requires recalculating waypoints at least every second while using only 20% of the CPU, so we had to trade off speed and complexity in both waypoint calculation and monitoring. The initial version of the UV EA detected the first five types of alerts listed in Section 7.4.

### 7.3 UV-Robotics: Execution Monitoring Issues

The initial monitoring issues apparent within the UV-Robotics domain can be divided into the following four categories:

- Monitoring the completion of, or progress toward, a basic action (e.g., go to a waypoint)

- Monitoring the satisfaction or completion of the multiple tasks to which the robot is currently committed (e.g., pursue evader, patrol area, photograph target every 2 hours)

- Monitoring the activity of unknown or adversarial entities

- Monitoring the state of the communication network, the robot, and other team members (e.g., communication network quality or integrity, robot mobility, or battery level)

Comparing these to our ontology in Section 4, the first two categories involve the general alert types *plan constraint violated* and *constraint violation projected*. However, they exist at different levels of abstraction and often have different temporal impact and associated monitoring requirements. The third category cleanly fits the *adversarial activity detected* alert type and triggers alerts for both autonomous control and user reporting. The fourth category is essential both to team-based automated operation and effective user interaction, and involves *policy constraint violated* alerts, *reporting requirement* alerts, and *system problem detected* alerts.

### 7.4 The UV-Robotics Execution Assistant

Like the SUO EA, our robot controller uses a rich plan representation to allow team members to share context and communicate with the user. Primitive actions in this domain are basic motion control and communication requests to the physical robot. A goal request from the user is decomposed into individual agent plans (recipes) and intentions to aid or interact with other agents. Recipes are composed of partially ordered sequences of tasks that in turn evolve into primitive actions.

The UV EA uses an internal architecture similar to the SUO EA, as shown in Figure 3. As in the SUO EA, the EA Manager continually applies its Acts to respond to new goals and facts posted in its database. The Acts correspond to algorithms for monitoring requirements at each layer in the MLAA architecture. Some implement user alerts and others implement autonomous control.

The inputs to the UV EA are plans to execute, policy declarations, status reports (including location, speed and orientation) from its own sensor suite, and messages from other agents. These messages include status reports of other agents, reports on mission success or failure, shared information, and requests for help. Depending on communication conditions or policy restrictions, an agent may, or may not, receive from team members status reports (up-to-date locations) of all friendly agents and other entities within visual range. Sentinels are extracted from plans and policy declarations, are evaluated when status reports are received, and may produce alerts. The alerts produced are designed to serve both the autonomous control via the plan manager component, and the user, although the needs of each vary considerably.

For our initial experimentation, all monitoring alerts were derived from regular *state* messages from each team member. A state message reports the current location, velocity, attitude, and sensor imprecision of an agent. A UV-Robotics Act similar to the SUO Act in Figure 5 is invoked every time a location report is posted in the EA Manager database. Such postings happen several times each second, because each robot receives two such messages every second from its own sensors and two from each team member, based on network conditions. It also receives similar state messages about entities within its own field of vision. This means in a team of three robots each agent will be handling a minimum of at least six state messages per second and possibly many more depending on the environment. Also, there are messages between agents for sharing information, which we are not currently considering except when they update state knowledge about adversarial entities. In the future, the UV EA will be extended to serve the higher layers of the architecture that have more in common with the SUO-EA alert types and triggers.

Our initial implementation of the UV EA detects the following types of alerts. We plan to implement additional monitoring during the project.

- **At-goal** – robot at current waypoint

- **Stuck** – robot stuck and not at current waypoint

- **Divergent** – robot diverging from current waypoint

- **No-status** – robot no longer reporting its state

- **Target-visible** – robot has a target within its sensor range

- **Lost-target** – robot lost track of target during pursue mission

- **Target-gone** – target moved out of assigned sector during pursue mission

- **Collision** – robot anticipates it will hit a nearby object in the next few seconds

- **Handoff** – robot has delegated/accepted a task to/from another team member

The UV EA uses the same techniques as the SUO EA (Section 6.7) for estimating VOA and greatly reducing the number of low-value alerts. In particular, the UV EA keeps event histories for each team member being monitored. These histories are used to determine the value of information and alerts, and to detect Stuck, Divergent, and No-status alerts. For example, the history indicates the time and the robot location at the last progress check, so if the current waypoint has not changed and the robot is further away from the waypoint, then the value of issuing a Divergent alert should be calculated.

The value of issuing an alert takes into consideration customizable latency thresholds and repetition parameters, which are associated with both the automated agent and the user. Some of the agent parameters are customized to improve performance, while others are a function of the behavior of the robot. For example, the value of a *divergent* alert will be a function of the expected velocity of the agent, because an agent traveling at speed will diverge more quickly than a slow agent. Similarly, a change in orientation will influence the value of an alert because a turning agent, while not decreasing the distance to the waypoint, may indeed be making progress toward the goal.

An example of how monitoring is used to facilitate autonomous control is illustrated by the situation where an agent is patrolling a designated area. When an evader becomes visible, the agent receives a *Target-visible* alert, which is of type *adversarial activity detected*. Reacting to either a high-priority policy to pursue evaders or an explicit plan step, the agent commits to a new goal *Pursue named-evader*. This goal is achieved by the activation and blending of three tasks: *Follow named-evader*, *Relocate named-evader* and *Search-for named-evader*. Thus, the robot will maintain pursuit even when the evader slips in and out of its field of vision.

The user's preferred strategy might be to report the first sighting of the evader or to track its position, noting whenever it disappears from view. However, the autonomous control requires notification only if the likelihood of recovering visual contact is deteriorating and the robot is searching aimlessly. At this point, a *Target-Lost* alert, which is of type *plan constraint violated*, will be sent to the agent's EA Manager (and possibly the user). In this example, a policy exists for reacting to this type of alert. It will cause the pursuit goal to be dropped and the original *Patrol* plan to be resumed.

## 7.5 UV-Robotics: Evaluation

The UV EA is being evaluated within an SRI experimental framework called the SRI Augmented Reaility Simulator (SARS) (Ortiz et al., 2002). The framework allows our autonomous agent architecture and software to be tested within an entirely simulated environment, on a team of physical robots, or a mixture of the two. The physical robots are three pioneer robots from equipped with GPS, as shown in Figure 10. Initial experiments were carried out in a simulated environment. We then ran the system in an entirely physical world with a team of two cooperating robots searching for and pursuing two independent evader robots. We have also run in environments composed of a combination of physical robots and simulated entities to illustrate scalability and operation with UAVs. The monitoring technology was effective in ensuring robust execution in all environments, and in giving human operators insight into the state and activity of each robot. This insight facilitated debugging and the process of moving from the simulated world to the physical robots as problems were quickly identified.

SARS is specifically designed to simulate robots and UAVs. It produces the same output in terms of sensors, actuators, and resources (battery status, communication range, and so forth). SARS computation and simulation is based on a precise 3D model of the environment. SARS is precise enough that we can mix physical robots moving in the real world with virtual evaders and see the physical robots following a virtual evader — thus, the name *augmented reality*. Using SARS, we are able to simulate a team of UGVs moving and/or UAVs flying in a larger space than we have available. The team of UAVs may be larger than our available physical UAVs, as well.

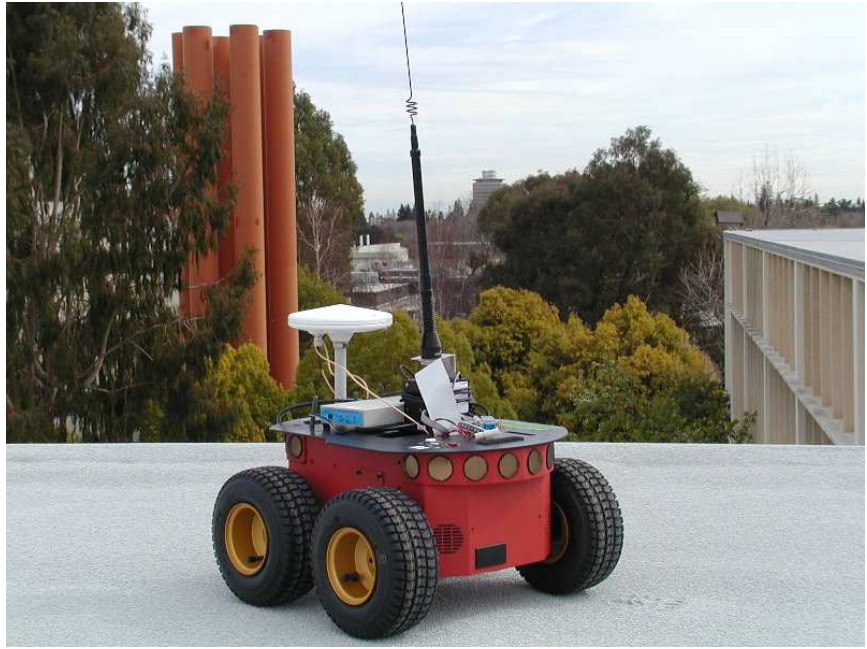

Figure 10: SRI experimental Pioneer UGV.

The initial UV EA implementation was evaluated with respect to the usefulness of its output, value of alerts, and real-time performance with realistic data streams. Our analysis shows that all the important situations are alerted during simulated executions, and during tests with actual robots, which are never exactly reproducible.

*No-status* alerts have proven useful to the human user, as they indicate a hardware or software problem on a robot or the network. Such problems are recognized immediately (after the customizable interval of noncommunication has passed) with the UV EA, but take considerably longer to detect without alerts from the EA. A customizable threshold (which currently defaults to 5 seconds) determines the value of an alert when the robot has not reported its state for a certain interval.

*At-goal*, *Stuck*, and *Divergent* are essential alerts for the autonomous-control agent-navigation system, as well as being useful to a human user who wants to monitor the activity of a single robot. Knowing when the robot has reached a goal point, when it has stopped and is not making progress toward a goal point, and when it is diverging from the planned route are essential to robust autonomous operation. Customizable intervals also control these alerts. Subtleties of the domain must be considered to avoid false alarms. For example, the robot may be paused because of GPS uncertainty and the GPS should be given time to establish connection with satellites. Also, a robot takes time to turn and thus should not be regarded as stuck or divergent until turns and steering adjustments have had time to complete.

*Target-visible*, *Target-lost*, and *Handoff* are useful to both the user and the autonomous controller, particularly when the task is to monitor or pursue a target. The autonomous controller requires immediate awareness of loss of sensor contact, so it can adjust its lower-level behavior or sensor parameters to find the evader. However, such immediate alerts would be unproductive for the human user or the plan-level controller. A customizable interval gives the agent time to relocate the evader, possibly avoiding an alert to the human. These types of alerts are the most time critical in our evaluation domain.

Good tracking of an evader requires recalculating waypoints and orientation at least every second. The UV EA was able to keep up with data inputs, detect occurrences of the types of alerts mentioned within 1 second, and recalculate waypoint and orientation twice per second. These constraints were not difficult to meet on our desktop machines, but the success of the UV EA on the slower processors of the physical robot involved tradeoffs of speed with complexity of waypoint calculation and monitoring. One useful technique is only using the latest state report for an agent when more than one state report has accumulated during one cycle of the monitoring loop. The relative CPU access of the various agents and processes also became important. For example, we had to adjust the time quantum given by the scheduler to our EA processes to ensure that both the process receiving messages and the various PRS agent processes in our EA were executed frequently enough for waypoint recalculation. This problem has been alleviated with more recent upgrades in the onboard computer, but could recur if more computationally expensive projections or alerts are added to the EA.

## 8. Related Work

Plan generation has received a lot of attention recently, but rarely are the plans used to control and monitor execution. Even more rarely are plans monitored that involve the activity of hundreds of agents requiring tight coordination. Previous work on execution monitoring has focused on models where the executor performs the planned actions (e.g., a robot controller) and usually has direct access to internal state information. In the SUO domain, most actions are performed by external agents, usually humans, and the monitor has no access to the state of its executing agents. Such indirect execution requires different monitoring techniques, as the executor must use incoming messages to determine the status of agents and activities and whether actions have been initiated or completed. The Continuous Planning and Execution Framework (Myers, 1999) has addressed the indirect execution problem, and our system builds on its ideas. However, our domain requires monitoring of many more constraints with greater time sensitivity. We have much higher rates of incoming data, and must customize monitoring of each action to generate appropriate, high-value alerts.

Robot designers have often avoided the plan representations used by the AI plan-generation community because of their restrictive assumptions (Pollack & McCarthy, 1999; Arkin, 1998). Both our domains required an expressive plan representation, and our combination of the Act formalism with a hierarchical, object-oriented mission model proved sufficiently expressive, providing a rich set of goal modalities for encoding activity, including notions of achievement, maintenance, testing, conclusion, and waiting.

The SAM system (Kaminka & Tambe, 1999) at ISI addresses a similar problem: automated pilot agents on a battlefield. SAM has direct access to its local automated agent and much lower incoming data rates than the EA. It addresses the difficult problem of plan recognition (of the plans of other friendly agents). Because humans are not involved, SAM does not need to produce alerts tailored to human cognitive capabilities. Experiments with SAM showed that distributed monitoring outperformed centralized monitoring while using simpler algorithms. Our EAs and SAIM use such a distributed design, building on these insights.

More recent work at ISI has produced a monitoring agent named OVERSEER (Kaminka et al., 2001), which also addresses a problem similar to ours: many geographically distributed team members with a coordinating plan in a dynamic environment. address the problem of modeling the value of information to the user. OVERSEER does not use the report-based monitoring approach adopted by our EAs, because it must rely on unmodifiable legacy agents and does not have sufficient bandwidth and reliability in communication. A detailed analysis is given in Section 3.

NASA's Remote Agent on Deep Space One (Jonsson et al., 2000; Muscettola et al., 1998) does autonomous execution monitoring on a spacecraft. Our domains have many of the same requirements as NASA's, including the core requirements of concurrent temporal processes and interacting recoveries. However, NASA's remote agent is fully automated, which places a heavier burden on the module that generates plans and responses, but alleviates the burden of having to address human interaction issues such as those considered in VOA. Monitoring algorithms are not described in detail, but are based on a procedural executive, which we assume is similar to our procedural reactive control system. In NASA's domain, the "agents" are mechanical devices onboard the spacecraft, and their behaviors have been formally modeled. Our agents include humans, whose behaviors are not easily modeled, so our EAs estimate the value of alerts as they interact with a human decision maker, who ultimately is responsible for the control decisions.

Work on rationale-based monitoring (Pollack & McCarthy, 1999; Veloso, Pollack, & Cox, 1998) addressed the problem of monitoring the world during the plan generation process (in causal-link planners) to see if events invalidate the plan being generated. They monitor subgoals, preconditions, usability conditions, and user preferences. All these are monitored in our framework when plans are executed, and our EAs have additional capabilities, such as monitoring policy constraints and applying mission-specific monitoring methods. This rationale-based work does not address time-critical monitoring during execution time, monitoring large volumes of incoming data, or the problem of alerting users without overwhelming them.

Doyle (1995) describes a technique to focus the user's attention on anomalous system behavior, particularly sensor behavior. This work would be applicable within the lowest layer in our robotics control module. It uses causal modeling to understand the "normal" behavior of a sensor. Anomaly detection is based on measures of *causal distance* and *distance* from normal behavior. The distance measures are not related to the plan and its goals/actions; instead they measure deviation from typical behavior. The user still has to relate the reported sensor anomaly to its higher-level effects, such as a threat to plan or action execution. This work provides a monitoring technique for specific sensor and system types that could easily be incorporated in our monitoring framework. The resulting anomaly detection might give low-level alerts or be a contributory factor in the reasoning process for higher-level alert classes.

The Phoenix system uses the concept of a plan envelope (Hart, Anderson, & Cohen, 1990) to represent the *a priori* expectations of an action's progress. Envelopes are used when an action executes over time and can be interrupted and altered during execution. The envelope captures the range of possible performance of an action during successful execution. During execution, the actual performance of the system is recorded and, if it deviates from the predefined envelope, a possible failure is detected. This concept provides a useful monitoring technique for specific alert-types, particularly those concerning actions that consume a variable amount of resources over time. Envelopes can also identify when an action is performing better than required allowing opportunistic alerts. Envelopes could easily be incorporated in our monitoring framework as an additional monitoring technique, and could be useful at the higher levels in both our domains.

The SUO EA provides a capability that does not currently exist, because there is no machine-understandable representation of the plan on the battlefield. Currently, small-unit warfighters must monitor all incoming information for relevance, with manual notification of other team members. The SUO EA also improves on next-generation Army systems such as FBCB2 (Force XXI Battle Command Brigade and Below) (Garamone, 2001). Unlike FBCB2, the EA alerts only on important changes, can automatically update the areas to be monitored as the plan is executed, can dynamically change the force structure, and can alert the user to many issues that are not monitored in other systems, such as fratricide risks, triggering of contingencies, and schedule, coordination and positional deviations from the plan.

## 9. Conclusions

We characterized the domain-independent challenges posed by an execution aid that interactively supports humans monitoring the activity of distributed teams of cooperating agents, both human and machine. The most important issues for interactive monitoring are adaptivity, plan- and situation-specific monitoring, reactivity, and high-value, user-appropriate alerts. We showed how properties of various domains influence these challenges and their solutions. We then presented a top-level domain-independent categorization of the types of alerts a plan-based monitoring system might issue to a user. The different monitoring techniques generally required for each category are often domain specific and task specific.

Our monitoring framework integrates these various techniques and then uses the concept of *value of an alert* to control interaction with the user. This conceptual framework facilitates integration of new monitoring techniques and provides a domain-independent context for future discussions of monitoring systems. We discussed various design tradeoffs that must be made during the application of our monitoring framework to a domain (Sections 6.4 and 6.6).

We use this framework to describe a monitoring approach we developed and have used to implement Execution Assistants (EAs) in two different dynamic, data-rich, real-world domains. Our approach is based on rich plan representations, which allow the execution aid to filter, interpret, and react to the large volume of incoming information, and alert the user appropriately. An expressive plan representation is necessary for representing SUO plans, which must coordinate distributed units, trigger contingencies, and enforce a variety of constraints. It is equally important that this representation be monitorable by machines and meaningful to humans. Our plan representation and mission model were able to model a representative SUO scenario with enough fidelity to provide value (as judged by our domain experts) and was also sufficient for plans in the UV-Robotics domain.

We developed a sufficiently rich plan representation by extending an existing plan representation with a hierarchical, object-oriented mission model that encodes knowledge about primitive actions and mission-specific monitoring methods. The SUO EA implements a novel integration of these hierarchical monitoring methods with a reactive control system. The EA invokes the most specific methods defined in the hierarchy at appropriate points during monitoring.

One central challenge, in our domains as well as medical monitoring, is to avoid overwhelming the user with unwanted alerts and/or false alarms. We define the concepts of value of information and value of giving an alert as the principles for determining when to give an alert. We describe the properties of VOI and VOA, criteria for computing them, the advantages of qualitative reasoning in our domains, and the successful use of these concepts in our applications. VOI and VOA algorithms must be customizable to the user, plan, and situation.

By using an asynchronous multiagent architecture and an extended version of the PRS reactive control system, we monitored the execution of both SUO and UV-Robotics plans with acceptable latency, given a dozen or more incoming events per second. PRS extensions include temporal monitors and efficiency improvements. Methods from the mission model are used throughout the SUO monitoring process for action-specific monitoring. Our evaluation showed that our plan-aware EAs generated appropriate alerts in a timely manner without overwhelming the user with too many alerts, although a small percentage of the alerts were unwanted. We have shown the utility of using advanced AI planning and execution technologies in small unit operations.

The application to UV-Robotics showed the generality of our SUO framework and monitoring concepts. We implemented a complex execution assistant in about one person-week, using code from the SUO EA. The UV EA uses the same plan representation and basic architecture as the SUO EA, but the inputs are different as are the tasks and the algorithms that respond to the inputs and generate alerts.

**Future work.** The most obvious area for future work in the SUO domain is incorporation of a planning assistant to complete the loop of continuous planning and execution. This integration has already been accomplished in the UV-Robotics domain, but the difficulty in the SUO domain is an interface that allows a soldier to interact effectively with the planning tool, using a wearable computer in a battlefield situation. Several research programs are addressing this problem, some of which are mentioned in Section 2.

Within the scope of execution monitoring, future work on our EAs could model and detect other types of plan deviations (such as loss of surprise or additional types of fratricide risks), project future failures, and provide higher-fidelity specialized reasoners, particularly for terrain reasoning. Additional theoretical work on VOI and VOA would support better quantitative estimates of VOI and VOA. The SUO mission model already has a method for projecting failures and a low-fidelity projection capability could be easily added. In the UV-Robotics domain, we plan to implement additional types of alerts in the near future, and extend the UV EA to serve the higher layers of the architecture that have more in common with the SUO EA alert types and triggers. The fragility of the UV communication network in hostile domains provides a set of interesting monitoring challenges that may result in the incorporation of specific monitoring-related tasks within cooperative team missions. Monitoring strategies for uncertain communication environments is an important research challenge for the UV-Robotics domain. Additional alerts being considered for future implementation include monitoring movement of entities in and out of geographical sectors mentioned in the plan, monitoring the deterioration or improvement of communication conditions, and monitoring the actions and intentions of coordinating team members to facilitate cooperative behavior.

## ACKNOWLEDGMENTS

The SUO research was supported by Contract F30602-95-C-0235 with the Defense Advanced Research Projects Agency (from the DARPA Planning and Decision Aids Program and the DARPA Small Unit Operations Program), under the supervision of Air Force Research Laboratory – Rome. The UCAV research was supported by the Office of Naval Research Unmanned Combat Air Vehicles Program (Contract N00014-00-C-0304). The SRI International Artificial Intelligence Center supported the writing of this paper. We thank the subject matter experts who assisted us. Our primary collaborators and evaluators were Kenneth Sharpe of SAIC and Richard Diehl of the Institute for Defense Analyses. We also used the expertise of Andy Fowles, Chris Kearns, and David Miller of the U.S. Army Dismounted Battlespace Battle Laboratory (DBBL) at Fort Benning, and CPT Dan Ray of the Mounted Maneuver Battlespace Laboratory (MMBL) at Fort Knox.

## References

Arkin, R. (1998). *Behavior-based robotics*. MIT Press.

Ash, D., Gold, G., Seiver, A., & Hayes-Roth, B. (1993). Guaranteeing real-time response with limited resources. *Artificial Intelligence in Medicine*, *5*(1), 49–66.

Athey, S., & Levin, J. (2001). The value of information in monotone decision problems. Tech. rep., Stanford University, Stanford, CA.

Bell, B., Jr., E. S., & Brown, S. M. (2002). Making adversary decision modeling tractable with intent inference and information fusion. In *Proc. of the 11th Conference on Computer Generated Forces and Behavioral Representation*, Orlando, FL.

Bonasso, R. P., Kortenkamp, D., & Whitney, T. (1997). Using a robot control architecture to automate space shuttle operations. In *Proc. of the 1997 National Conference on Artificial Intelligence*, pp. 949–956, Providence, RI. AAAI Press.

Cheyer, A., & Martin, D. (2001). The open agent architecture. *Journal of Autonomous Agents and Multi-Agent Systems*, *4*(1), 143–148.

Coiera, E. (1993). Intelligent monitoring and control of dynamic physiological systems. *Artificial Intelligence in Medicine*, *5*(1), 1–8.

Donlon, J., & Forbus, K. (1999). Using a geographic information system for qualitative spatial reasoning about trafficability. In *Proc. of the Qualitative Reasoning Workshop*, Loch Awe, Scotland.

Doyle, R. J. (1995). Determining the loci of anomalies using minimal causal models. In *Proc. of the 1995 International Joint Conference on Artificial Intelligence*, pp. 1821–1827, Montreal, Quebec, Canada. Morgan Kaufmann Publishers Inc., San Francisco, CA.

Durfee, E. H., Huber, M. J., Kurnow, M., & Lee, J. (1997). TAIPE: Tactical assistants for interaction planning and execution. In *Proc. of Autonomous Agents '97*. ACM Press, New York.

Ferguson, G., & Allen, J. (1998). TRIPS: An integrated intelligent problem-solving assistant. In *Proc. of the 1998 National Conference on Artificial Intelligence*, pp. 567–572. AAAI Press.

Forbus, K. D. (2002). Towards Qualitative Modeling of the Battlespace. Technical report unpublished manuscript, Northwestern University, Evanston, IL.

Franke, J., Brown, S. M., Bell, B., & Mendenhall, H. (2000). Enhancing teamwork through team-level intent inference. In *Proc. of the 2000 International Conference on Artificial Intelligence*, Las Vegas, NV.

Garamone, J. (2001). Digital world meets combat during desert exercise. Tech. rep., American Forces Information Service, www.defenselink.mil/news/Apr2001/.

Georgeff, M. P., & Ingrand, F. F. (1989). Decision-making in an embedded reasoning system. In *Proc. of the 1989 International Joint Conference on AI*, pp. 972–978, Detroit, MI. Morgan Kaufmann Publishers Inc., San Francisco, CA.

Gil, Y., & Blythe, J. (1999). A problem-solving method for plan evaluation and critiquing. In *Proc. of the Tenth Banff Knowledge Acquisition for Knowledge-Based Systems Workshop*, Banff, Alberta, Canada.

Grosz, B., & Kraus, S. (1999). The evolution of SharedPlans. In Rao, A., & Wooldridge, M. (Eds.), *Foundations and Theories of Rational Agencies*, pp. 227–262.

Hart, D. M., Anderson, S. D., & Cohen, P. R. (1990). Envelopes as a vehicle for improving the efficiency of plan execution. Tech. rep. UM-CS-1990-021, University of Massachusetts, Amherst, MA.

Holness, G., Karuppiah, D.and Uppala, S., & Ravela, S. C. (2001). A service paradigm for reconfigurable agents. In *Proc. of the 2nd Workshop on Infrastructure for Agents, MAS, and Scalable MAS (Agents 2001)*, Montreal, Canada.

Horty, J., & Pollack, M. (2001). Evaluating new options in the context of existing plans. *Artificial Intelligence*, *127*(2), 199–220.

Jonsson, A., Morris, P., Muscettola, N., & Rajan, K. (2000). Planning in interplanetary space: Theory and practice. In *Proc. of the 2000 International Conference on AI Planning and Scheduling*, pp. 177–186, Breckenridge, CO. AAAI Press, Menlo Park, CA.

Kaminka, G., Pynadath, D., & Tambe, M. (2001). Monitoring deployed agent teams. In *Proc. of Autonomous Agents '01*, pp. 308–315, Montreal, Canada.

Kaminka, G., & Tambe, M. (1999). Experiments in distributed and centralized socially attentive monitoring. In *Proc. of Autonomous Agents '99*, pp. 213–220, Seattle, WA.

Konolige, K., & Myers, K. (1998). *Artificial Intelligence Based Mobile Robots: Case studies of Successful Robot Systems*, chap. The Saphira architecture: a design for autonomy. MIT Press.

Koski, E., Makivirta, A., Sukuvaara, T., & Kari, A. (1990). Frequency and reliability of alarms in the monitoring of cardiac postoperative patients. *International Journal of Clinical Monitoring and Computing*, *7*, 129–133.

Mouaddib, A.-I., & Zilberstein, S. (1995). Knowledge-based anytime computation. In *Proc. of the 1995 International Joint Conference on Artificial Intelligence*, pp. 775–783. Morgan Kaufmann Publishers Inc., San Francisco, CA.

Muscettola, N., Nayak, P. P., Pell, B., & Williams, B. C. (1998). Remote agent: To boldly go where no AI system has gone before. *Artificial Intelligence*, *103*(1-2), 5–47.

Musliner, D. J., Durfee, E. H., & Shin, K. G. (1993). CIRCA: A cooperative intelligent real-time control architecture. *IEEE Transactions on Systems, Man, and Cybernetics*, *23*(6).

Myers, K. L. (1996). A procedural knowledge approach to task-level control. In *Proc. of the 1996 International Conference on AI Planning Systems*. AAAI Press, Menlo Park, CA.

Myers, K. L., & Morley, D. N. (2001). Human directability of agents. In *Proc. 1st International Conference on Knowledge Capture*, Victoria, B.C.

Myers, K. L. (1999). CPEF: A continuous planning and execution framework. *AI Magazine*, *20*, 63–70.

Ortiz, C., Agno, A., Berry, P., & Vincent, R. (2002). Multilevel adaptation in teams of unmanned air and ground vehicles. In *First AIAA Unmanned Aerospace Vehicles, Systems, Technologies and Operations Conference*.

Ortiz, C. L. (1999). Introspective and elaborative processes in rational agents. *Annals of Mathematics and Artificial Intelligence*, *25*(1–2), 1–34.

Ortiz, C. L., & Hsu, E. (2002). Structured negotiation. In *Proc. of the First International Conference on Autonomous Agents and Multiagent Systems*.

Pollack, M. E., & McCarthy, C. (1999). Towards focused plan monitoring: A technique and an application to mobile robots. In *Proc. of the IEEE International Symposium on Computational Intelligence in Robotics and Automation (CIRA)*, pp. 144–149.

Rao, A. S., & Georgeff, M. P. (1995). BDI-agents: From theory to practice. In *Proc. of the First Intl. Conference on Multiagent Systems*, San Francisco.

Schreckenghost, D., & et al. (2001). Adjustable control autonomy for anomaly response in space-based life support systems. In *Proc. of the IJCAI Workshop on Autonomy, Delegation, and Control*.

Shannon, C. (1948). A mathematical theory of communication. *Bell System Technical Journal*, *27*, 379–423, 623–656.

Tsien, C. (1997). Reducing false alarms in the intensive care unit: A systematic comparison of four algorithms. In *Proc. of the American Medical Informatics Association Annual Fall Symposium*.

Tsien, C., & Fackler, J. (1997). Poor prognosis for existing monitors in the intensive care unit. *Critical Care Medicine*, *25*(4), 614–619.

Veloso, M., Pollack, M., & Cox, M. (1998). Rationale-based monitoring for planning in dynamic environments. In *Proc. of the 1998 International Conference on AI Planning Systems*, pp. 171–180. AAAI Press, Menlo Park, CA.

Vincent, R., Horling, B., Lesser, V., & Wagner, T. (2001). Implementing soft real-time agent control. In *Proceedings of the 5th International Conference on Autonomous Agents*. ACM Press.

Weigner, M. B., & Englund, C. E. (1990). Ergonomic and human factors affecting anesthetic vigilance and monitoring performance in the operating room environment. *Anesthesiology*, *73*(5), 995–1021.

Weinberger, E. (2002). A theory of pragmatic information and its application to the quasispecies model of biological evolution. *Biosystems*, *66*(3), 105–119.

Wilkins, D. E., & desJardins, M. (2001). A call for knowledge-based planning. *AI Magazine*, *22*(1), 99–115.

Wilkins, D. E., & Myers, K. L. (1995). A common knowledge representation for plan generation and reactive execution. *Journal of Logic and Computation*, *5*(6), 731–761.

Wilkins, D. E., & Myers, K. L. (1998). A multiagent planning architecture. In *Proc. of the 1998 International Conference on AI Planning Systems*, pp. 154–162, Pittsburgh, PA.

Wilkins, D. E., Myers, K. L., Lowrance, J. D., & Wesley, L. P. (1995). Planning and reacting in uncertain and dynamic environments. *Journal of Experimental and Theoretical AI*, *7*(1), 121–152.